\documentclass[12pt,reqno]{amsart}
\usepackage{amsmath}
\usepackage{amsxtra}
\usepackage{amscd}
\usepackage{amsthm}
\usepackage{amsfonts}
\usepackage{amssymb}
\usepackage{eucal}
\usepackage{cases}
\usepackage{paralist} 
\usepackage{array}
 \usepackage{etex} 
\usepackage{color}
\usepackage{float}
\usepackage{cite}

\usepackage{graphicx,color}

\allowdisplaybreaks

\def\smallskip{\vskip\smallskipamount}
\def\medskip{\vskip\medskipamount}
\def\bigskip{\vskip\bigskipamount}

\def\({\left(}
\def\){\right)}
\def\[{\left[}
\def\]{\right]}
\newcommand{\ds}[1]{\displaystyle #1}

\newcommand{\nn}{\nonumber}
\newcommand{\bea}{\begin{eqnarray}}
	\newcommand{\ena}{\end{eqnarray}}
\def\bel{\begin{eqnarray}}
	\def\enl{\end{eqnarray}}
\newcommand{\be}{\begin{eqnarray*}}
	\newcommand{\en}{\end{eqnarray*}}
\newcommand{\ba}{\begin{array}}
	\newcommand{\ea}{\end{array}}

\newcommand{\slt}{\mathfrak{sl}_2}
\newcommand{\sltr}{\mathfrak{sl}_3}
\newcommand{\slth}{\widehat{\mathfrak{sl}}_2}
\newcommand{\slthree}{\widehat{\mathfrak{sl}}_3}

\renewcommand{\Im}{\mathop{\rm Im}}
\newenvironment{tenumerate}{
	\begin{enumerate}
		
	}{\end{enumerate}}
\newcommand{\bi}{\begin{tenumerate}}
	\newcommand{\ei}{\end{tenumerate}}
\newcommand{\isoto}[1][]%
{{\mathop{\buildrel{\sim}\over\longrightarrow}\limits_{#1}}}

 \let\be=\beta

  \let\la=\lambda 
    
\let\om=\omega

\def\2{\frac{1}{2}} \def\4{\frac{1}{4}}

\def\6{\partial}

\def\+{\dagger}

\def\<{\langle} \def\>{\rangle}

\def\ctg{\, {\rm ctg}\,}

\renewcommand{\cot}{\ctg}

\def\Re{{\rm Re\,}} \def\Im{{\rm Im\,}}

\numberwithin{equation}{section}

\begin{document}
	
\begin{title}[Calculating the correlation functions of the rational $\sltr$-model]
	{On the calculation of the correlation functions of the $\sltr$-model
	 by means of the reduced qKZ equation}
\end{title}
\date{\today}
\author{H.~Boos, A.~Hutsalyuk, Kh.~S.~Nirov}
\address{Physics Department, University of Wuppertal, D-42097,
	Wuppertal, Germany}\email{hboos@uni-wuppertal.de}
\address{Physics Department, University of Wuppertal, D-42097,
	Wuppertal, Germany}
\address{  Moscow Institute of Physics and Technology,  Dolgoprudny, Moscow reg., Russia}\email{hutsalyuk@gmail.com}
\address{Institute for Nuclear Research of the Russian Academy of Sciences,
	60th October Ave 7a, 117312 Moscow, Russian Federation,}\email{nirov@inr.ac.ru}
\address{International Laboratory of Representation Theory and Mathematical Physics, National Research University Higher 
School of Economics, Moscow, Russian Federation}\email{hnirov@hse.ru}

\begin{abstract}
We study the reduced density matrix of the $\sltr$-invariant fundamental
exchange model by means of a novel reduced quantum Knizhnik-Zamolodchikov
equation. This gives us insight into the algebraic structure and explicit
results for correlation functions in the infinite chain ranging over up
to three sites.
\end{abstract}

\maketitle

\section{Introduction}

Over the past three decades considerable progress has been made in the
study of correlation functions of integrable models, in particular, of
the integrable spin-1/2 Heisenberg-Ising (or XXZ) chain associated with
the affine quantum group $U_q(\slth)$. Due to this
progress we are nowadays able to compute the static correlation functions of this
model under very general equilibrium conditions at short and large
distances and with arbitrary numerical accuracy \cite{BDGKSW08,TGK10a,%
SABGKTT11,DiFraSmi,MiwaSmi,KKMST09a}. What may be considered even more important
is that we may have uncovered part of the mathematical structure that
seems to distinguish the correlation functions of integrable systems
from those of non-integrable ones. As far as the mathematical structure
and the short-range correlation functions are concerned it turned out
to be particularly fruitful to study the reduced density matrix of
a finite connected chain segment of the XXZ chain and its inhomogeneous
multi-spectral-parameter version associated with the underlying
six-vertex model. Starting with the pioneering work \cite{Jimboetal} of
the `Kyoto school' a number of rather diverse methods have been applied
in order to derive various representations of the reduced density
matrix as well as general theorems about its structure. In
\cite{Jimboetal} a multiple-integral representation for the ground-state
correlation functions of the XXZ chain in the massive regime was
obtained by means of a `q-vertex operator approach' based on the
representation theory of $U_q(\slth)$ and inspired by Baxter's corner
transfer matrix method \cite{Baxter76a} and by conformal field theory
\cite{BPZ84}. A generalization to the massless regime was obtained in
\cite{JiMi96} using a different method based on functional equations in
the spectral parameters of qKZ-type \cite{FreResh,Smi92}. These
results were further generalized, utilizing the Bethe Ansatz, to include
a longitudinal magnetic field \cite{Mailletetal} and arbitrary finite
temperatures \cite{Goehmannetal}.


In papers \cite{BoKor}, it was observed that the multiple integrals representing
the ground-state correlation functions at vanishing magnetic field factorize in
products of one-dimensional integrals, this way reproducing a singular and
puzzling result of Takahashi \cite{Takahashi77} from the 70s. In collaboration of one
of the authors with M. Jimbo, T. Miwa, F. Smirnov and Y. Takeyama the
algebraic structure behind the factorization of the multiple integrals was
eventually unveiled \cite{HGS,HGSII}. Creation and annihilation operators on a
space of quasi-local operators were constructed in such a way that the
creation operators generate a special `fermionic basis' by iterated action
on an appropriate Fock vacuum \cite{HGSII,BJMS09a}. Jimbo, Miwa and Smirnov
\cite{HGSIII} then proved an important theorem showing that, under very general
conditions, the correlation functions corresponding to any quasi-local operator
can be related to only two functions $\rho$ and $\omega$ through a determinant
formula.
The function $\rho$ is a spectral-parameter dependent one-point function,
basically equal to the magnetization, while $\omega$ is a two-point function
depending on two spectral parameters. It can be interpreted as an expectation
value of a special operator of length 2 or a special nearest-neighbour
two-point correlation function. In this sense $\omega$ is very similar to the
energy density. Thus, the theorem of Jimbo, Miwa and Smirnov states that the
two most local non-trivial, independent correlation functions determine all
others through algebraic relations. Moreover all physical parameters, like
system length, temperature or magnetic field enter only through these two
functions which turn out to have efficient descriptions in terms of solutions
of linear and non-linear integral equations \cite{BoGo09}.


It would appear natural if a similar structure, implying that all static
correlation functions are algebraically determined by a few short-range
correlation functions, would exist for other, more complicated integrable
systems as well. Arguably, next in complexity, after the basic models
related to $U_q(\slth)$, come their higher-rank generalizations or the
rational counterparts of these models. Extending and elaborating the
works \cite{HibbKhorBazh,Tsuboi}, two of the authors studied the
functional equations related to their spectrum by means of representation
theory and gave their full proof in a universal form
in joint work with G{\"o}hmann, Kl{\"u}mper and Razumov \cite{BGKNR14}.

%

Studies of the spectral problem of these higher rank models, especially in
their rational, $\mathfrak{sl}(N)$-symmetric version have a long history
starting with the pioneering works by Yang \cite{Yang} and Sutherland \cite{Suth}
on the multi-component Bose and Fermi gases. In paper \cite{Suth75}, Sutherland
considered the diagonalization problem of the Hamiltonian 
\bea
\label{sutherlandham}
H=\pm \sum\limits_{j=1}^L P_{j,j+1}
\label{Ham}
\ena
with periodic boundary conditions, where the operator $P$ permutes 
local states of $N$ Bosons and $M$ Fermions. Nowadays we would call this
Hamiltonain the $\mathfrak{sl}(N|M)$-invariant exchange Hamiltonian. Sutherland
was the first to diagonalize it by nested Bethe Ansatz. A nested algebraic
Bethe ansatz was later developed by Kulish and Reshetikhin \cite{Resh,Kulish85}.
The $R$-matrix which is relevant in the $\mathfrak{sl} (N)$-case is proportional to 
the $S$-matrix of the $SU(N)$ Gross-Neveu model, which is an interesting
quantum field theory associated with a higher rank quantum group
\cite{GrossNeveu,Witten,Zam-Zam,BabuKarowski}.

The correlation functions of relativistic quantum field theories
can be studied starting from solutions to a set of functional equations
known as the `form factor axioms' \cite{Karowskietal,book}. The form
factors give access to correlation functions through certain spectral
representations alias sums over multiple integrals which are often
useful for asymptotic analysis. Form factors for lattice models like
(\ref{sutherlandham}) are less constrained. They have not been obtained
as solutions of functional equations, but in recent years have been
constructed by algebraic Bethe ansatz methods \cite{Slavnovetal,PRS14}. It seems
however difficult to sum them up to correlation functions, especially
at short-distances. The only more concrete result for correlation
functions that has been obtained so far by means of the lattice form-factor
approach concerns the two-point functions at large distances \cite{KoRa16}.

In papers \cite{Koyama,Kojima} the vertex-operator approach was used to
obtain the multiple integral representation for the $U_q(\hat{\mathfrak{sl}}_n)$
model in critical and massive regimes. Unfortunately, the formulas are rather
bulky and are given only up to the normalization. This certainly makes 
a precise numerical computation difficult.

It is moreover unclear if an algebraic structure, similar
to the one identified for the XXZ model, exists for the fundamental
$\mathfrak{sl}_3$ lattice model.

In this paper we consider the rational $\sltr$-model with Hamiltonian (\ref{sutherlandham})
at zero temperature and zero external fields. Since we were unable to directly generalize
the hidden fermionic structure of the $\slt$-case we shall follow the original
idea that led to its discovery. We shall derive a reduced quantum Knizhnik-Zamolodchikov
equation (rqKZ) introduced for the $\slt$-case in \cite{BJMT040506}. Then we solve the novel
rqKZ equation for $n=1,2,3$ lattice sites, giving us direct analytical results for
the reduced density matrices of chain segments of the respective length.
Note that the rqKZ equation was not known for the $\sltr$-case and that,
as we shall see, a generalization from $\slt$ to $\sltr$ is not as
straightforward as it might appear, since there is no crossing symmetry
in the higher-rank case. Our results for the short-range correlation
functions for $n = 2, 3$ are, to the best of our knowledge, the first
explict results for correlation functions of this model. The case $n=4$
is considerably more complicated. We will describe it in a separate
publication. The purpose and the driving force behind our work is the
hope to identify a minimal number of independent short-range
correlation functions that will determine all correlation functions
of the model at larger distances.

The plan of the paper is as follows. In Section 2, we describe the
integrable structure of the $\sltr$-model. We introduce the corresponding
$R$-matrix and recall some of its properties, such as unitarity and crossing
relations. In Section 3, we discuss static correlation functions, related
density matrices, inhomogeneous generalizations and their basic properties.
Section 4 is the main section of this work. Here we introduce a pair of reduced
qKZ equations and the resulting closed rqKZ equation for the generalized
density matrix which we solve for three lengths of the corresponding local
operators, $n=1,2,3$. To solve this equation, we introduce two transcendental
functions $\om^{(1)}$ and $\om^{(3)}$ and discuss their properties. We also
discuss the homogeneous limit of the formulas we have obtained for the density
matrices. Section 5 is devoted to conclusions. In Appendix \ref{s:apa}, we
provide a heuristic derivation of our rqKZ equation based on a technique 
developed by Aufgebauer and Kl\"umper for the $\slt$-case at finite
temperature \cite{AufKlum}. In Appendix \ref{s:apb}, we show some details
of the derivation of the integral representation of our function  $\om^{(3)}$
in homogeneous case.

\section{The rational $\sltr$ model}

Let us start with the formulation of the model. In the more general case, including a deformation parameter $q$, 
the $R$-matrix is defined in the tensor
product of two representation spaces of $U_q(\slthree)$. The Khoroshkin--Tolstoy formula \cite{TolKho92} 
for the universal $R$-matrix 
allows one to obtain the $R$-matrix for arbitrary 
representations in the so-called auxiliary and quantum spaces \cite{BGKNR10}, 
but we are firstly interested in the case of two fundamental representations. 
Taking the limit $q \to 1$ in the expression for the $U_q(\slthree)$-related $R$-matrix from 
\cite{BGKNR10}, we reproduce the $R$-matrix of the rational $\sltr$-model under consideration.  
It has a particularly simple form 
\bea
\ds{{R_{12}(\la)}= \frac{\rho(\la)}{\la+1}\bigl(\la\;\mathbf{1}_{12}+P_{12}\bigr)},
\label{R}
\ena
where $\mathbf{1}_{12}$ and $P_{12}$ are the $3^2 \times 3^2$ unit and permutation
matrices, respectively, acting in the tensor product of spaces $1$ and $2$ in the case 
of three states or ``colors'', and the function{\footnote{Note that this function 
$\rho$ is not the function $\rho$ mentioned in the Introduction. We hope that the 
reader will not be confused about it.}}
$\rho$ can be regarded as a quasi-classical, $q \to 1$, or the rational, limit of 
the transcendental pre-factor obtained from the Kho\-rosh\-kin--\-Tols\-toy formula. 
In our case, it satisfies two functional relations
\bea
\rho(\la)\rho(-\la)=1, \qquad 
\rho(\la)\rho(3-\la)=\frac{(\la-3)\;\la}{(\la-1)(\la-2)}, 
\label{functilderho}
\ena
an appropriate solution of which can explicitly be written in the form
\bea
\ds{\rho(\la)=-\frac{\Gamma(\la/3)\Gamma(1/3-\la/3)}{\Gamma(-\la/3)\Gamma(1/3+\la/3)}}.
\label{tilderho}
\ena 
Alternatively, $\rho(\lambda)$ can be obtained using the algebraic structure 
of the Yangian double of $\sltr(\mathbb{C})$ \cite{Smi92,KhoTol94}. 

Also we need to define matrices $\bar R$ and $\bar{\bar R}$ 
acting in the tensor product of fundamental and anti-fundamental representations 
and in the tensor product of anti-fundamental and fundamental representations, respectively. 
To this end, we use the corresponding crossing relations
\bea
&{\bar R}_{1\bar{2}}(\la)=C_{\bar 2, 2} {\bigl(R_{12}(-\la-1)\bigr)}^{t_2}C_{2,\bar 2},\nn\\
&{\bar{\bar R}}_{\bar{1}2}(\la)=C_{\bar 1, 1} {\bigl(R_{12}(-\la-2)\bigr)}^{t_1}C_{1,\bar 1},
\label{barR}
\ena
where $t_j$ denotes the transposition in space $j$ and $C_{\bar j, j}=C_{j,\bar j}$ stand for the ``charge conjugation'' 
matrix $C$ 
\bea
C=\begin{pmatrix}
	0 & 0 & 1\\
	0 & 1 & 0\\
	1 & 0 & 0
\end{pmatrix}.
\label{C}
\ena
Below we will use $C_{i,\bar j}=C_{\bar j,i}$ for incoming lines $i$ and $\bar j$ and 
$C^{i,\bar j}=C^{\bar j,i}$ for the outgoing lines $i$ and $\bar j$. In both cases it 
is again given by (\ref{C}). 

When it does not cause any misunderstanding, we will just write $R_{i,j}$ for $R_{i,j}(\la_{i,j})$, where 
$\la_{i,j}=\la_i-\la_j$ and similarly for the other $R$-matrices $\bar R$ and $\bar{\bar R}$. We can also explicitly 
write down 
\bea
&&{\bar R}(\la)=\frac{\bar{\rho}(\la)}{\la}\bigl(\la+1-C\otimes C\bigr),
\quad {\bar\rho}(\la)={\rho}(\la+1)^{-1},
\label{barRbarbarR}\\
&&{\bar {\bar R}}(\la)={\bar R}(\la+1).
\nn
\ena
An important immediate consequence of this formula and definition (\ref{R}) is that{\footnote{ 
Here we take into account that $\lim_{\varepsilon \to +0} \frac{\Gamma(\varepsilon)}{\Gamma(-\varepsilon)} = - 1$.}} 
\bea
R(0)=P,\quad {\bar R}(-1)={\bar {\bar R}}(-2)=C\otimes C.
\label{RbarRspecial}
\ena

For later usage we will also need four relations which are consequences of the 
Yang--Baxter equation and the above crossing relations 
(\ref{barR}) and (\ref{RbarRspecial}):
\bea
&& R_{1,3}(\la_{1,3})\,{\bar{\bar R}}_{\bar 2,3}(\la_{1,3}-2)\,C_{1,\bar 2}=
\mathbf{1}_3\; C_{1,\bar 2},
\quad
{\bar R}_{1,\bar 3}(\la_{1,3})\,R_{\bar 2,\bar 3}(\la_{1,3}-2)\,C_{1,\bar 2}=\mathbf{1}_{\bar 3}\; C_{1,\bar 2},
\label{cross1}\\
&& {\bar{\bar R}}_{\bar 1,3}(\la_{1,3})\, R_{2,3}(\la_{1,3}-1)\,C_{\bar 1, 2}=
	\mathbf{1}_3\; C_{\bar 1, 2},
\quad
R_{\bar 1,\bar 3}(\la_{1,3})\,{\bar R}_{2,\bar 3}(\la_{1,3}-1)\,C_{\bar 1, 2}=\mathbf{1}_{\bar 3}\; C_{\bar 1, 2}.
\label{cross2}
\ena
It is not difficult to verify these relations directly.

\section{Static correlation functions and  density matrix}

If one takes all external fields to be zero, the quasi-local operators become just local 
operators.\footnote{A local operator is an operator localized on a finite fraction of the 
lattice, while a quasi-local operator is the product of such a local operator with a factor (`a tail') 
having a simple dependence on an external disorder field; for this and other related notions we refer 
to papers \cite{HGS,HGSII}} In this case the static zero temperature correlation function 
of some local operator $X_{1,\ldots,n}$ defined on a lattice segment of the length $n$ is 
the vacuum expectation value 
\bea
\langle\text{vac}|X_{1,\ldots,n}|\text{vac}\rangle =
D\bigl(X_{1,\ldots,n}\bigr),
\label{cor}
\ena
where $|\text{vac}\rangle$ corresponds to the ground state of the model in the thermodynamic limit. 
$D$ stands for the density matrix\footnote{This concept is naturally understood in the same sense 
as in the framework of quantum mechanics.} acting on a local operator $X$. As was discussed in 
papers \cite{HGSII,HGSIII}, it is a functional which maps any operator $X_{1,\ldots,n}$ to a number.
The elements of the density matrix are defined as follows:
\bea
D_{i_1,\cdots,i_n}^{i'_1,\ldots,i'_n}=D(E_{i_1}^{i'_1}\otimes\cdots\otimes E_{i_n}^{i'_n}),
\label{Delements}
\ena
where  $E_{i}^{i'}$ are the elements of the basis of $\mathrm{Mat}_3(\mathbb{C})$ 
corresponding to the standard basis of $\mathbb{C}^3$, and so, these are the standard matrix 
units. We will also use the shorthand notation $D_{1,\ldots,n}$. 

A useful trick is to introduce the inhomogeneity parameters $\la_1, \ldots, \la_n$
for the above segment of length $n$. The new ground state will depend
on these parameters
$$
|\text{vac}\rangle\to |\text{vac}_{\{\la_1,\ldots,\la_n\}}\rangle.
$$
One can define a generalized  density matrix which
also depends on  $\la_1, \ldots, \la_n$:
\bea
\langle\text{vac}_{\{\la_1,\ldots,\la_n\}}|
X_{1,\ldots,n}|\text{vac}_{\{\la_1,\ldots,\la_n\}}\rangle =
D(\la_1,\ldots,\la_n)\bigl(X_{1,\ldots,n}\bigr),
\label{corinhom}
\ena
with the corresponding matrix elements
$D_{1,\ldots,n}(\la_1,\ldots,\la_n)$ defined in the same way as in (\ref{Delements}).
The generalized density matrix shows much more structure. We will see that this information 
can help us to find an explicit solution for the generalized density 
matrix. After such a solution is found, one can obtain 
the original density matrix by taking the homogeneous limit 
\bea 
D_{1,\ldots,n}=\lim_{\la_1\rightarrow 0, \ldots, \la_n\rightarrow 0 }D_{1,\ldots,n}(\la_1, \ldots, \la_n).
\label{homolim}
\ena

Let us list some important properties of the generalized density matrix.

\begin{compactenum}[(i)]	
\item
The normalization condition 
\bea 
D(\la_1,\ldots,\la_n)(\mathbf{1}_{1,\ldots,n})=1 
\label{normD}
\ena
is consistent with the reduction relations.
	
\item
Left--right reduction relations 
	\bea
	&& D(\la_1,\ldots,\la_n)(\mathbf{1}_1X_{2,\ldots,n})=D(\la_2,\ldots,\la_n)(X_{2,\ldots,n}),\nn\\ && D(\la_1,\ldots,\la_n)(X_{1,\ldots,n-1}\mathbf{1}_n)=D(\la_1,\ldots,\la_{n-1})(X_{1,\ldots,n-1}).
	\label{leftrightred}
	\ena

\item
The asymptotic condition
	\bea
	\lim_{\la_1\to\infty}D_{1,\ldots,n}(\la_1,\ldots, \la_n)=\frac13\; \mathbf{1}_1\;D_{2,\ldots,n}(\la_2, \ldots, \la_n).
	\label{asymp}
	\ena 
		
\item 
The $R$-matrix relations
	\bea
	D_{1,\ldots,i,i+1,\ldots,n}(\la_1,\ldots, \la_n)\bigl(R_{i,i+1}(\la_{i,i+1})X_{1,\ldots,n}R_{i+1,i}(\la_{i+1,i})\bigr)
	\nn\\
	= D_{1,\ldots,i+1,i,\ldots, n}(\la_1,\ldots,\la_{i+1},\la_{i},\ldots \la_n)\bigl(X_{1,\ldots,n}\bigr).
	\label{RmatD}
	\ena
	
	
\item
The translational invariance 
\bea
 D(\la_1+u,\ldots ,\la_n+u)=D(\la_1,\ldots ,\la_n)
\label{translinv}
\ena
implying that the generalized density matrix elements depend only on differences
$\la_i-\la_j$.
	

\item
The global ${GL}_3$-invariance
\bea
G\otimes\cdots\otimes G\;\Bigl(D(\la_1,\ldots,\la_n)\Bigr)\;
G^{-1}\otimes\cdots\otimes G^{-1}=D(\la_1,\ldots,\la_n),
\label{GL3}
\ena
where $G$ is any element of the ${GL}_3(\mathbb{C})$ group in the fundamental representation.

\item	
The color conservation
	\bea
	&& D(\la_1,\ldots, \la_n)_{i_1,\ldots,i_n}^{i'_1,\ldots,i'_n}\ne 0,
	\quad \text{only if}\quad  n_1(\{i\})= n_1(\{i'\}),\nn
	\\
	&&\quad n_2(\{i\})= n_2(\{i'\}),
	\quad n_3(\{i\})= n_3(\{i'\}), \label{colconserve}
	\ena
	where $n_j(\{i\})$ is the number of indices of ``color'' $j$ in the $n$-tuple  
	$\{i\}=\{i_1,\ldots, i_n\}$. There is a symmetry with respect to permutations 
	of colors.
	
\end{compactenum}	

	
	%
	%
	%
	%
	
	
The above properties (i)--(v) are rather similar to the corresponding properties in the $\slt$-case. 
Therefore, we will not prove them here. The  properties (vi), (vii) follow directly 
from  the characteristics of the $R$-matrix (\ref{R}).

\section{The rqKZ equations for the rational $\sltr$ model}

Our experience with the $\slt$-case suggests that the above 
properties do not fix the correlation functions uniquely. The missing information is hidden
in a set of additional equations of difference type which were called the reduced qKZ equation 
\cite{BJMT040506}. We need to deduce such equations for the $\sltr$-case as well. This is done 
heuristically in Appendix \ref{s:apa}, where we obtain the following pair of  difference 
equations:{\footnote{We learned about the existence of these equations first in a seminar 
talk at Wuppertal University given by G.P.A.~Ribeiro in February 2017.}
\begin{align}
& D 
(\la_1,\la_2,\ldots,\la_n)\Bigl(A^{(1)}_{1,\bar 1|2,\ldots,n}
(\la_1|\la_2,\ldots,\la_n)
\bigl(X_{1,2,\ldots,n}\bigr)\Bigr)=
D^{(1)} 
(\la_1-2,\la_2,\ldots,\la_n)\bigl(X_{\bar 1,2,\ldots,n}\bigr),
\label{rqKZ1} \\
& D^{(1)} 
(\la_1,\la_2,\ldots,\la_n)
\Bigl(A^{(2)}_{\bar 1,1|2,\ldots,n}(\la_1|\la_2,\ldots,\la_n)
\bigl(X_{\bar 1,2,\ldots,n}\bigr)\Bigr)=\
D 
(\la_1-1,\la_2,\ldots,\la_n)\bigl(X_{1,2,\ldots,n}\bigr).
\label{rqKZ2}
\end{align}
Here the density matrix $D^{(1)}$ describes the situation 
with one anti-fundamental representation in the first quantum space 
and fundamental representations associated with the other spaces $2\cdots n$
(as depicted in Fig.~4 of Appendix \ref{s:apa} in the more general case).
The operator $A^{(1)}_{1,\bar{1}|2,\cdots,n}$ is defined as follows: 
it acts on some local operator $X_{\bar{1},2,\cdots,n}$ 
as a matrix ${\Bigl({A^{(1)}}_a^b\Bigr)}_{\bar{1}|2,\cdots,n}$ with 
respect to the space 1 with incoming line $a$ and outgoing line $b$
\begin{multline}
{\bigl({A^{(1)}}_a^b\bigr)}_{\bar{1}|2,\ldots,n}(\la_1|\la_2,\ldots,\la_n)
\bigl(X_{\bar{1},2,\ldots,n}\bigr) \\
:= C^{\bar{1},b}R_{2,b}(\la_{2,1})\cdots R_{n,b}(\la_{n,1})X_{\bar{1},2,\ldots,n}
R_{b,n}(\la_{1,n}) \cdots R_{b,2}(\la_{1,2})C_{\bar{1},a}, 
\label{A1}
\end{multline}
and the operator $A^{(2)}_{\bar 1, 1|2,\ldots,n}$  
acts on some local operator $X_{1,\ldots,n}$ 
as a matrix 
with respect to the space $\bar{1}$ with incoming 
line $\bar{a}$ and outgoing line $\bar{b}$
\begin{multline}
{\bigl({A^{(2)}}_{\bar{a}}^{\bar{b}}\bigr)}_{1|2,\ldots,n}(\la_1|\la_2,\ldots,\la_n)
\bigl(X_{1, 2, \ldots,n}\bigr) \\
:= C^{1,\bar{b}} {\bar{R}}_{2,\bar{b}}(\la_{2,1}) \cdots {\bar{R}}_{n,\bar{b}}(\la_{n,1})
X_{1, 2, \ldots, n}
{\bar{\bar R}}_{\bar b,n}(\la_{1,n}) \cdots {\bar{\bar R}}_{\bar b,n}(\la_{1,2})C_{1,\bar a}.
\label{A2}
\end{multline}
In Appendix \ref{s:apa} we also show the above equations graphically in Fig.~\ref{qkz1} 
and Fig.~\ref{qkz2}.

Combining equations (\ref{rqKZ1}) and (\ref{rqKZ2}), we come to a novel closed reduced qKZ equation 
(rqKZ) which will be the key relation for solving the problem of the calculation 
of correlation functions in case of the $\sltr$-invariant model:
	\bea
	& D 
	(\la_1, \ldots, \la_n)\Bigl(A_{1|2,\ldots,n}(\la_1|\la_2, \ldots, \la_n)
	\bigl(X_{1,2,\ldots,n}\bigr)\Bigr)=
	D 
	(\la_1-3, \ldots, \la_n)\bigl(X_{1,2,\ldots,n}\bigr).
	\label{rqkz}
	\ena
Here  
\bea
A(\la_1|\la_2,\ldots,\la_n)=
A^{(1)}(\la_1|\la_2,\ldots,\la_n) 
A^{(2)}(\la_1-2|\la_2,\ldots,\la_n)
\label{operatorA}
\ena	
by definition.
Since this formula looks a bit formal, let us
explicitly write down the action of this operator on some 
local operator $X$. It acts with respect to the first space
as a matrix ${(A_a^b)}_{2,\ldots,n}(\la_1|\la_2,\ldots,\la_n)$
with incoming line $a$ and outgoing line $b$:
\begin{multline}
{(A_a^b)}_{2,\ldots,n}(\la_1|\la_2,\ldots,\la_n)(X_{1,\ldots,n}) \\
= C^{\bar 1,b}R_{2,b}(\la_{2,1}) \cdots R_{n,b}(\la_{n,1})
C^{1,\bar a}{\bar R}_{2,\bar a}(\la_{2,1}+2) \cdots 
{\bar R}_{n,\bar a}(\la_{n,1}+2)\bigl(X_{1,\ldots,n}\bigr) \\ 
\times {\bar{\bar R}}_{\bar a,n}(\la_{1,n}-2) \cdots {\bar{\bar R}}_{\bar a,2}(\la_{1,2}-2)
R_{b,n}(\la_{1,n}) \cdots R_{b,2}(\la_{1,2}) C_{\bar a,a}C_{1,\bar 1}.
\label{A}
\end{multline}
The corresponding picture is given in Appendix \ref{s:apa} (see Fig.~\ref{rqkzfinal}).

So far, we have been able to solve the whole set of 
relations (\ref{leftrightred})--(\ref{GL3}) together with the 
rqKZ equation (\ref{rqkz}) up to the length $n=3$. We present our results below.

\subsection{The case $n=1$.}

From the above symmetry and normalization (\ref{normD}) we immediately 
come to the conclusion that 
\bea
{D(\la_1)}_i^j=\frac13\delta_i^j.
\label{D1}
\ena

The rqKZ relation (\ref{rqkz}) gives in this case the following simple
equation:
\bea
{D(\la_1)}_i^j={D(\la_1-3)}_i^j,
\label{rqkz1}
\ena
which is compatible with solution (\ref{D1}). 
One can easily check that all other relations 
(\ref{leftrightred})--(\ref{GL3}) are 
fulfilled automatically.

\subsection{The case $n=2$.}

The case $n=2$ is more substantial. From the 
color conservation property (\ref{colconserve}) we conclude that 
there are only three non-trivial non-zero elements  
\bea
&& D_0(\la_1,\la_2)={D(\la_1,\la_2)}_{i,i}^{i,i}\quad \text{for} \quad i=1,2,3,
\label{D2}
\\
&& D_1(\la_1,\la_2)={D(\la_1,\la_2)}_{i,j}^{i,j},\quad 
D_2(\la_1,\la_2)={D(\la_1,\la_2)}_{i,j}^{j,i}\quad 
\text{for} \quad i,j=1,2,3
\quad \text{and}\quad i\ne j.
\nn
\ena
From the global $GL_3$-invariance (\ref{GL3}) it follows  
that the density matrix is of the form 
\bea
D_{1,2}(\la_1,\la_2)=\mathbf{1}_{1,2}\,D_1(\la_1,\la_2)+
P_{1,2}\,D_2(\la_1,\la_2).
\label{DfromGL3}
\ena
Certainly, the subscripts 1 and 2 of the functions $D_1,D_2$ in this formula 
should not be mixed up with the numbers of spaces 1 and 2.

Thus, we can immediately obtain the functions 
$D_0$ and $D_1$ from the reduction relations 
(\ref{leftrightred}). We have  
\bea
D_0(\la_1,\la_2)=D_1(\la_1,\la_2)+D_2(\la_1,\la_2)=\frac13-2 D_1(\la_1,\la_2),
\label{D0}
\ena
and so, 
\bea
D_1(\la_1,\la_2)=\frac19-\frac13 D_2(\la_1,\la_2).
\label{D1a}
\ena 
From the $R$-matrix invariance (\ref{RmatD}) and translational invariance (\ref{translinv}) 
it follows that the function $D_2$ is symmetric
\bea
&& D_2(\la_1,\la_2)=D_2(\la_2,\la_1),\quad 
\label{symfunn2}
\ena
and depends on the difference of the spectral parameters. Let us choose it
in the following form:
\bea
D_2(\la_1,\la_2)=-3\,\om^{(1)}(\la_{1,2}),
\label{om}
\ena 
where $\om^{(1)}(\la)$ is an even function of $\la$.
With this choice formula (\ref{DfromGL3}) 
becomes
\bea
D_{1,2}(\la_1,\la_2)=\frac19\mathbf{1}_{1,2}+
\om^{(1)}(\la_{1,2})\,P^{(0)}_{1,2},
\label{D12sol}
\ena
where the notation 
\bea 
P^{(0)}_{1,2}=\mathbf{1}_{1,2}-3P_{1,2}
\label{P0P1}
\ena
is introduced.

 
Now we have to solve the rqKZ relation.
To this end, we can substitute formula (\ref{D12sol}) 
into the rqKZ relation  (\ref{rqkz}) and note that 
\bea
&& P^{(0)}_{1,2}A_{1|2}(\la_1|\la_2)=P^{(1)}(\la_{1,2}) P^{(0)}_{1,2},
\label{P0A}\\
&&  \mathbf{1}_{1,2} A_{1|2}(\la_1|\la_2)=\mathbf{1}_{1,2}+9\,Q^{(1)}(\la_{1,2})P^{(0)}_{1,2},
\label{P1A}
\ena
where $P^{(1)},Q^{(1)}$ are  rational functions
\bea
P^{(1)}(\la)=\frac{(\la-4)(\la-2)}{(\la-1)(\la+1)},\quad 
Q^{(1)}(\la)=-\frac{2\la-3}{3(\la-3)(\la-1)\la(\la+1)}.
\label{PQ}
\ena
We see that the operator $P^{(0)}$ turns out to be 
a constant eigenvector of the operator $A$ with the eigenvalue 
$P^{(1)}(\la)$. Two relations (\ref{P0A}), (\ref{P1A}) 
are in fact nothing but the reduction of the rqKZ equation to the triangular 
form. Therefore, if we act on $A$ by the right-hand side of (\ref{D12sol}), 
use formula (\ref{P0A}), (\ref{P1A}) and equate 
the coefficients standing before the identity operator and the operator $P^{(0)}$, 
we come to the following functional relation for the function $\om^{(1)}$:
\bea
\om^{(1)}(\la-3)=P^{(1)}(\la)\om^{(1)}(\la)+Q^{(1)}(\la).
\label{ommin3}
\ena
This equation is nothing but the result of diagonalization of the rqKZ equation, 
which is related to a certain non-local matrix Riemann--Hilbert problem. It is not 
clear yet how to find its solution in general case. 

It is interesting to note that the coefficients $P^{(1)}$ and $Q^{(1)}$ should satisfy 
certain compatibility condition
\bea
P^{(1)}(3-\la)P^{(1)}(\la)=1,\quad P^{(1)}(3-\la)Q^{(1)}(\la)+Q^{(1)}(3-\la)=0,
\label{compatibility1}
\ena
which might be seen as a zero curvature condition in some geometric picture. 
We will consider this question in more detail elsewhere.

The solution to the above functional relation (\ref{ommin3}) looks as
\bea
&&\om^{(1)}(\la)=-\frac{1}{36}+\frac{\la^2-1}{24}\bar{\om}^{(1)}(\la),
\label{om1}
\ena
where the function
\bea 
\bar{\om}^{(1)}(\la)=-\frac{\partial}{\partial\la}\log{\rho(\la)}
\label{barom1a}
\ena
satisfies the functional equation
\bea
\bar{\om}^{(1)}(\la-3)=\bar{\om}^{(1)}(\la)+\bar{Q}^{(1)}(\la)
\label{barom1func}
\ena 	
with 
\bea
\bar{Q}^{(1)}(\la)=\frac1{\la-3}-\frac1{\la-2}-\frac1{\la-1}+\frac1{\la}=
\frac{2(2\la-3)}{(\la-3)(\la-2)(\la-1)\la}.
\label{barQ1}
\ena
For later use we will need two further representations of the function $\bar{\om}^{(1)}(\la)$:
\bea
\bar{\om}^{(1)}(\la) \!\!\! &=& \!\!\! -\frac{1}{3}\Bigl(
\psi\Bigl(\frac{\la}3\Bigr)+\psi\Bigl(-\frac{\la}3\Bigr)
-\psi\Bigl(\frac13+\frac{\la}3\Bigr)-\psi\Bigl(\frac13-\frac{\la}3\Bigr)\Bigr) 
\label{barom1b} \\
&=& \!\!\! -\sum_{j=0}^{\infty}\Bigl(\frac1{\la-3j}-\frac1{\la+3j}
+\frac1{\la+1+3j}-\frac1{\la-1-3j}\Bigr),
\label{barom1c}
\ena
where $\psi(\la)$ is the logarithmic derivative of the $\Gamma$-function.
Bellow we will also use $\psi_n(\la)=\partial_{\la}^n\psi(\la)$.

Thus, we obtain the entries of the generalized density matrix 
\bea
&& {D(\la_1,\la_2)}_{i,i}^{i,i}=D_0(\la_1,\la_2)=\frac19-2\,\om^{(1)}(\la_{1,2}),\nn\\
&& {D(\la_1,\la_2)}_{i,j}^{i,j}=D_1(\la_1,\la_2)=\frac19+\om^{(1)}(\la_{1,2}),
\label{Dres}\\
&& {D(\la_1,\la_2)}_{i,j}^{j,i}=D_2(\la_1,\la_2)=-3\,\om^{(1)}(\la_{1,2}),\nn
\ena
where it is implied that $i \ne j$ in last two equations.

Since for large values of $|\la|$ we have
\bea
\bar{\om}^{(1)}(\la) \simeq \frac{2}{3} \la^{-2} - \frac{2}{3} \la^{-4} + \mathcal{O}(\la^{-6}),
\nn
\ena
the asymptotic behavior of the function $\om^{(1)}(\la)$ is as follows:{\footnote{Strictly speaking, 
one should take $\la=u+iv$ with a finite real part $-3<u<3$ and send $v\to\pm\infty$, but we can also set 
$\la\to\la_0+3m$ with some real $\la_0$ close to $0$ and then take the limit for an integer $m\to\pm\infty$. 
Below we will always imply such a limit and just write $\la\to\infty$.}
\bea
&&
\om^{(1)}(\la)\simeq -\frac1{18}\, \la^{-2}+\mathcal{O}(\la^{-4})\quad{\mbox{ when}}\quad
{|\la|\to \infty}.
\label{asympom1}
\ena
Using this formula, we can easily check the asymptotic relation (\ref{asymp}).
It means that our result (\ref{Dres}) fulfills the rqKZ relation (\ref{rqkz})
and all the above properties (i)--(vii).

Now it is easy to obtain the elements of the original density matrix taking the 
homogeneous limit (\ref{homolim}), since the value of the function $\om^{(1)}(\la)$ 
at $\la=0$ is well defined{\footnote{Here we take into account that 
$\lim_{z \to 0} (\psi(z)+\psi(-z)) = -2\gamma$.}
\bea
\ds{\om^{(1)}(0)=
	\lim_{\la\to 0}\om^{(1)}(\la)=-\frac1{36}\Bigl(1+\gamma+\psi\bigl(\frac13\bigr)\Bigr)},
\label{om1at0}
\ena
where $\gamma$ is Euler's constant. 

Finally, the result for the density matrix in the $n=2$ case looks as follows:
\bea
&& {\ds D_{i,i}^{i,i}=\frac19-2\om^{(1)}(0)=
	\frac16+\frac1{18}\Bigl(\gamma+\psi\bigl(\frac13\bigr)\Bigr)=
	\frac16-\frac{\pi}{36\sqrt{3}}-\frac{\log{3}}{12}
		=0.0247323...}\nn\\
&& {\ds D_{i,j}^{i,j}=\frac19+\om^{(1)}(0)=\frac1{12}-\frac1{36}\Bigl(\gamma+\psi\bigl(\frac13\bigr)\Bigr)
	=\frac1{12}+\frac{\pi}{72\sqrt{3}}+\frac{\log{3}}{24}
	=0.154301...}\label{Dhomres}\\
&& {\ds D_{i,j}^{j,i}=-3\,\om^{(1)}(0)=
	\frac1{12}\Bigl(1+\gamma+\psi\bigl(\frac13\bigr)\Bigr)
	=\frac1{12}-\frac{\pi}{24\sqrt{3}}-\frac{\log{3}}{8}=
	-0.129568...}\nn
\ena
where we imply that $i,j=1,2,3$ and $i\ne j$.

Let us compare this with our numerical result obtained by direct diagonalization
of the transfer matrix up to the lengths $L=9,12$ (see Table 1)
\begin{table}[h]
	\begin{center}
		\setlength\extrarowheight{7.0pt}
		\begin{tabular}{|c|c|c|}
			\hline
			\qquad &\qquad $D_{1,2}^{1,2}$ \qquad &\qquad $D_{2,1}^{1,2}$ \\ 
			\hline
			Exact result ($L=\infty$) & 0.1543 & -0.129568\\
			\hline
			\qquad L=9\qquad& 0.15546 & -0.133048\\
			\hline
			\qquad L=12\qquad& 0.154946 & -0.131505 \\
			\hline
		\end{tabular}
	\end{center}
	\caption{\label{tablenum2} Comparison of numerical and analytic results for $D_2$}
\end{table}


\subsection{The case $n=3$.}

In a sense, the above result for the density matrix for $n=2$ is rather
similar to the $\slt$-case, where the function $\om$ there was related to the 
logarithmic derivative of the pre-factor of the $R$-matrix \cite{BJMT040506}.
We will see that the situation with the $n=3$ case is essentially different
since it will be necessary to involve one more function of three spectral 
parameters. Technically, the case $n=3$ is more intricate, and 
some formulas become rather tedious. Therefore, let us only roughly describe 
our basic steps that we made in order to come to the final result. 
\begin{compactenum}[(I)]	
\item
The first step is to use the global $GL_3$-invariance (\ref{GL3})
in order to write the elements of the generalized density matrix 
in the following form:
\bea
{D(\la_1,\la_2,\la_3)}_{i_1,i_2,i_3}^{i'_1,i'_2,i'_3}=\sum_{\sigma\in \mathcal{S}_3}
g^{(\sigma)}(\la_1,\la_2,\la_3)\; \delta_{i_1}^{i'_{\sigma(1)}}\delta_{i_2}^{i'_{\sigma(2)}}
\delta_{i_3}^{i'_{\sigma(3)}} \label{D3fromGL3},
\ena
where the sum goes over all six elements $\sigma$ of the permutation group $\mathcal{S}_3$
with six unknown functions  $g^{(\sigma)}(\la_1,\la_2,\la_3)$.
\item
The second step is to fulfill the $R$-matrix relations  (\ref{RmatD}) 
together with the reduction relations (\ref{leftrightred}). 
At this stage we need
to solve some functional relations in order to express the above six unknown  
functions $g^{(\sigma)}(\la_1,\la_2,\la_3)$ in terms of five 
fully symmetric functions $\tilde{g}^{(i)}(\la_1,\la_2,\la_3)$
with  $i=1,\cdots, 5$.
\item
The third step is to solve the rqKZ equation (\ref{rqkz}) with respect 
to these five functions. At this stage we observe that the final answer is a sum of
three terms: the first one is proportional to the identity operator which reduces by 
relations (\ref{leftrightred}) to the identity operator in the $n=2$ 
case in (\ref{Dres}), the second term reduces to terms  
containing the function $\om^{(1)}$, and the third term has zero reduction both from 
the left and from the right. This third term is proportional to a new fully symmetric 
function $\om^{(3)}(\la_1,\la_2,\la_3)$. With the help of the rqKZ equation (\ref{rqkz}) 
we find that the function $\om^{(3)}$ should fulfill certain functional 
relation which we discuss below.
\end{compactenum}
	
\vspace{0.5cm}

Let us show the result:
\begin{multline}
D_{1,2,3}(\la_1,\la_2,\la_3)=\frac{1}{27}\mathbf{1}_{1,2,3} \\
+ f_{1,2,3}^{(12)}(\la_1,\la_2|\la_3)\;\om^{(1)}(\la_{1,2}) 
+ f_{1,2,3}^{(13)}(\la_1,\la_2,\la_3)\;\om^{(1)}(\la_{1,3}) 
+ f_{1,2,3}^{(23)}(\la_1|\la_2,\la_3)\;\om^{(1)}(\la_{2,3}) \\
+ f_{1,2,3}^{(123)}\;\om^{(3)}(\la_1,\la_2,\la_3),
\label{resD3}
\end{multline}
where $f^{(ij)}$ are matrices with rational elements
\begin{multline}
f_{1,2,3}^{(12)}(\la_1,\la_2|\la_3) =
R_{1,2}\;f_{2,1,3}^{(12)}(\la_2,\la_1|\la_3)\;R_{1,2}^{-1}
\\
= \Bigl(1-\frac1{\la_{1,3}\la_{2,3}}\Bigr)\Bigl(\frac13-P_{1,2}\Bigr)+
\frac1{\la_{1,3}\la_{2,3}}P_{2,3} +
\frac12\Bigl(\frac{\la_{1,2}-3}{\la_{1,3}\la_{2,3}}P_{2,3}P_{1,2}
-\frac{\la_{1,2}+3}{\la_{1,3}\la_{2,3}}P_{1,2}P_{2,3}\Bigr), \label{f12} 
\end{multline}
\begin{multline}
f_{1,2,3}^{(13)}(\la_1,\la_2,\la_3)
= R_{2,3}\;f_{1,3,2}^{(12)}(\la_1,\la_3|\la_2)\;R_{2,3}^{-1} \\
= \Bigl(1-\frac1{\la_{2,1}\la_{2,3}}\Bigr)\Bigl(\frac13-P_{1,3}\Bigr)
+ \frac1{\la_{2,1}\la_{2,3}}(P_{1,2}+P_{2,3}-P_{1,3}) \\
+ \frac12\Bigl(\frac{\la_{2,1}+\la_{2,3}-3}{\la_{2,1}\la_{2,3}}
P_{1,2}P_{2,3}
-\frac{\la_{2,1}+\la_{2,3}+3}{\la_{2,1}\la_{2,3}}P_{2,3}P_{1,2}\Bigr),
\label{f13} 
\end{multline}
\begin{multline}
f_{1,2,3}^{(23)}(\la_1|\la_2,\la_3)
= R_{2,3}\;f_{1,3,2}^{(23)}(\la_1|\la_3,\la_2)\;R_{2,3}^{-1}
= R_{1,2}\;f_{2,1,3}^{(13)}(\la_2,\la_1,\la_3)\;R_{1,2}^{-1} \\
= \Bigl(1-\frac1{\la_{1,2}\la_{1,3}}\Bigr)\Bigl(\frac13-P_{2,3}\Bigr)
+ \frac1{\la_{1,2}\la_{1,3}}P_{1,2} +
\frac12\Bigl(\frac{\la_{2,3}-3}{\la_{1,2}\la_{1,3}}P_{1,2}P_{2,3}
-\frac{\la_{2,3}+3}{\la_{1,2}\la_{1,3}}P_{2,3}P_{1,2}\Bigr).
\label{f23}
\end{multline}
Here we did not write the identity operators in order not to overload the
formulas.

The coefficient $f^{(123)}$ in the last term in (\ref{resD3}) can be
written in two ways
\bea
f^{(123)}_{1,2,3}=
\frac{16}{9}\Bigl(1-\frac32(P_{1,2}+P_{1,3})\Bigr)\Bigl(1-\frac32 P_{2,3}\Bigr)
=
\frac{16}{9}\Bigl(1-\frac32 P_{1,2}\Bigr)\Bigl(1-\frac32(P_{1,3}+P_{2,3})\Bigr),
\label{f123}
\ena
where the zero left reduction is evident from the first formula
and the zero right reduction is evident from the second one. Also one
can check that (\ref{f123}) satisfies the $R$-matrix relations
\bea
f_{1,2,3}^{(123)}=R_{1,2}\;f_{2,1,3}^{(123)}\;R_{1,2}^{-1}
=R_{2,3}\;f_{1,3,2}^{(123)}\;R_{2,3}^{-1}.
\label{Rf123}
\ena
Note that, as in the case $n=2$, the coefficient $f^{(123)}$ is the 
eigenvector of the $A$-operator which does not depend on the spectral 
parameters:
\bea
f^{(123)}_{1,2,3}A_{1|23}(\la_1|\la_2,\la_3)=
P^{(3)}(\la_1|\la_2,\la_3)\; f^{(123)}_{1,2,3},
\label{f123A}
\ena
where the rational function 
$P^{(3)}(\la_1|\la_2,\la_3)$ is defined below.

As we announced above, the function $\om^{(3)}(\la_1,\la_2,\la_3)$ is
symmetric with respect to its arguments and satisfies the
functional relation
\begin{multline}
\om^{(3)}(\la_1-3,\la_2,\la_3)
= P^{(3)}(\la_1|\la_2,\la_3)\;\om^{(3)}(\la_1,\la_2,\la_3) \\
+ P^{(3,1)}(\la_1|\la_2|\la_3)\;\om^{(1)}(\la_{12})	
+P^{(3,1)}(\la_1|\la_3|\la_2)\;\om^{(1)}(\la_{13})
+P^{(3,2)}(\la_1|\la_2,\la_3)\;\om^{(1)}(\la_{23}) \\
+ Q^{(3)}(\la_1|\la_2,\la_3)
\label{om3func}
\end{multline}
with the rational functions				
\bea
&& P^{(3)}(\la_1|\la_2,\la_3)=\frac{(\la_{1,2}-4)(\la_{1,2}-2)\la_{1,2}}
{(\la_{1,2}-3)(\la_{1,2}-1)(\la_{1,2}+1)}
\cdot\frac{(\la_{1,3}-4)(\la_{1,3}-2)\la_{1,3}}
{(\la_{1,3}-3)(\la_{1,3}-1)(\la_{1,3}+1)},
\label{P3}\\
&& P^{(3,1)}(\la_1|\la_2|\la_3)=\frac{9(\la_{1,2}-4)(\la_{1,2}-2)\la_{1,2}}
{8(\la_{1,2}-1)(\la_{1,2}+1)(\la_{1,3}-3)(\la_{1,3}-1)(\la_{1,3}+1)\la_{2,3}},
\label{P31}\\
&& P^{(3,2)}(\la_1|\la_2,\la_3)=
-\frac{9(\la_{1,2}+\la_{1,3}-3)(\la_{2,3}^2-9)}{8(\la_{1,2}-3)
(\la_{1,2}-1)(\la_{1,2}+1)(\la_{1,3}-3)(\la_{1,3}-1)(\la_{1,3}+1)},
\label{P32}\\
&& Q^{(3)}(\la_1|\la_2,\la_3)=
-\frac{\la_{1,2}+\la_{1,3}-6}{8(\la_{1,2}-3)(\la_{1,2}-1)
(\la_{1,2}+1)(\la_{1,3}-3)(\la_{1,3}-1)(\la_{1,3}+1)}
\label{Q3}.
\ena
As in the $n=2$ case, the functions $P^{(3)},P^{(3,1)},P^{(3,2)}$ and $Q^{(3)}$ must satisfy
a set of compatibility conditions coming from the equation $\om^{(3)}(\la_1-3,\la_2-3,\la_3-3)=
\om^{(3)}(\la_1,\la_2,\la_3)$ and consequent application of the functional relation 
(\ref{om3func}). Let us show only one of them:
\bea
P^{(3)}(\la_1|\la_2-3,\la_3-3)P^{(3)}(\la_2|\la_1,\la_3-3)P^{(3)}(\la_3|\la_1,\la_2)=1.
\label{compatibilityn3}		
\ena 		
		
In fact, we have thus reduced the non-local matrix 
RH-problem (\ref{normD})--(\ref{colconserve}), 	(\ref{rqkz}) for $n=2,3$ to a 
non-local one-dimensional RH-problem.

		
Going backwards, it is not difficult to check that expression (\ref{resD3}) 
satisfies the rqKZ equation (\ref{rqkz}). It also satisfies the $R$-matrix relation 
(\ref{RmatD}) and the left and right reduction relations (\ref{leftrightred}). 

Of course, the solution of the difference relation (\ref{om3func}) is not unique. 
From the asymptotic condition (\ref{asymp}) we can conclude that the function 
$\om^{(3)}$ must decrease as some power of spectral parameters when they become large.
It fixes the unique solution. 
Our analysis of the asymptotic behavior at large spectral parameters
shows that 
\bea
\om^{(3)}(\la_1,\la_2,\la_3)\simeq \la_3^{-4}
\Bigl(-\frac{\la_{1,2}^2-9}{8}\;\om^{(1)}(\la_{1,2})-\frac1{144}\Bigr)+\mathcal{O}(\la_3^{-5})
\quad {\mbox{ when}}\quad \la_3\to\infty.
\label{asympom3}
\ena
Now, the asymptotic condition (\ref{asymp}) for $n=3$ can be checked using (\ref{asympom1})
and (\ref{asympom3}). This proves that expression (\ref{resD3}) really satisfies 
the whole set of properties (i)--(vii) from Section 3 and the reduced qKZ relation (\ref{rqkz}).

\subsection{Homogeneous limit}

In order to obtain the elements of the original density matrix (\ref{Delements}), we need to 
take the homogeneous limit $\la_i\to 0, i=1,2,3$ of the expression (\ref{resD3}). To this end we should calculate 
\bea
\om^{(3)}(\vec 0):=\lim_{\la_1\to 0,\la_2\to 0,\la_3\to 0} \om^{(3)}(\la_1,\la_2,\la_3).
\label{limom3}
\ena
We do it in three steps. 

First, we introduce the function 
\bea
\bar{\omega}^{(3)}(\la_1,\la_2,\la_3)=\prod_{j=2}^3 \frac{\la_{1,j}}{(\la_{1,j}-1)(\la_{1,j}+1)}
\om^{(3)}(\la_1,\la_2,\la_3),
\label{barom3}
\ena
which satisfies the functional relation
\begin{multline}
\bar{\omega}^{(3)}(\la_1-3,\la_2,\la_3)=\bar{\omega}^{(3)}(\la_1,\la_2,\la_3) \\
+\bar{P}^{(3,1)}(\la_1|\la_2|\la_3)\bar{\om}^{(1)}(\la_{1,2})
+\bar{P}^{(3,1)}(\la_1|\la_3|\la_2)\bar{\om}^{(1)}(\la_{1,3})
+\bar{P}^{(3,2)}(\la_1|\la_2,\la_3)\bar{\om}^{(1)}(\la_{2,3}) \\
+\bar{Q}^{(3)}(\la_1|\la_2|\la_3),
\label{barom3func}
\end{multline}
where the function $\bar{\om}^{(1)}$ was defined in (\ref{barom1a})--(\ref{barom1c}),
and the functions $\bar{P}$ and $\bar{Q}$ explicitly are 
\begin{align}
& \bar{P}^{(3,1)}(\la_1|\la_2|\la_3) = \frac{3}{64} \cdot 
\frac{(\la_{1,2}-3)\la_{1,2}}{(\la_{1,3}-4)(\la_{1,3}-2)(\la^2_{1,3}-1)\la_{2,3}},
\label{barP31} \\
& \bar{P}^{(3,2)}(\la_1|\la_2,\la_3)= -\frac{3}{64} \cdot 
\frac{(\la_{1,2}+\la_{1,3}-3)(\la_{2,3}^2-9)(\la_{2,3}^2-1)}
{(\la_{1,2}-4)(\la_{1,2}-2)(\la^2_{1,2}-1)(\la_{1,3}-4)(\la_{1,3}-2)(\la^2_{1,3}-1)},
\label{barP32}\\
& \bar{Q}^{(3)}(\la_1|\la_2,\la_3)= \frac{1}{32} \cdot 
\frac{(\la_{1,2}+\la_{1,3})(\la_{1,2}^2+\la_{1,3}^2+\la_{2,3}^2+13)-
	3(\la_{1,2}\la_{1,3}+4 \la_{1,2}^2+4\la_{1,3}^2)+27}
	{(\la_{1,2}-4)(\la_{1,2}-2)(\la^2_{1,2}-1)(\la_{1,3}-4)(\la_{1,3}-2)(\la^2_{1,3}-1)}. 
	\label{barQ}
\end{align}
From definition (\ref{barom3}) and the asymptotic behavior (\ref{asympom3}) 
we deduce that
\bea
\bar{\omega}^{(3)}(\la_1,\la_2,\la_3)\simeq \mathcal{O}(\la_1^{-6})\quad
{\mbox{at}}\quad \la_1\to\infty.
\label{asympbarom}
\ena

Our second step is to take the limit with respect to $\la_2,\la_3$:
\bea
\bar{\bar{\om}}^{(3)}(\la):=\bar{\om}^{(3)}(\la,0,0).
\label{barbarom}
\ena
From the functional relation (\ref{barom3func}) and definitions 
(\ref{barP31})--(\ref{barQ}) we deduce that the function 
$\bar{\bar{\om}}^{(3)}$ must satisfy the following relation:
\bea
\bar{\bar{\om}}^{(3)}(\la-3)-\bar{\bar{\om}}^{(3)}(\la)=r(\la),
\label{barbaromfunc}
\ena
where
\bea
r(\la)=p^{(3,1)}_1(\la)\bar{\om}^{(1)}(\la)+p^{(3,1)}_2(\la)\partial_{\la}\bar{\om}^{(1)}(\la)+
p^{(3,2)}(\la)\bar{\om}^{(1)}(0)+q^{(3)}(\la)
\label{r}
\ena
and
\bea
&&
p^{(3,1)}_1(\la)=-\frac3{64}\cdot\frac{(2 \la-3)(3\la^4-18\la^3+23\la^2+12\la-8)}{(\la-4)^2(\la-2)^2(\la-1)^2(\la+1)^2}\label{p31},\\
&&
p^{(3,1)}_2(\la)=-\frac3{64}\cdot\frac{( \la-3)\la}{(\la-4)(\la-2)(\la-1)(\la+1)}\label{dp31},\\
&&
p^{(3,2)}(\la)=-\frac{27}{64}\cdot\frac{2 \la-3}{(\la-4)^2(\la-2)^2(\la-1)^2(\la+1)^2}\label{p32},\\
&&
q^{(3)}(\la)=\frac1{32}\cdot\frac{4\la^3-27\la^2+26\la+27}{(\la-4)^2(\la-2)^2(\la-1)^2(\la+1)^2},\label{q3}
\ena
and from (\ref{asympbarom}) it follows that 
\bea
\bar{\bar{\omega}}^{(3)}(\la)\simeq \mathcal{O}(\la^{-6})\quad
{\mbox{at}}\quad \la\to\infty.
\label{asympbarbarom}
\ena

Using the functional relation (\ref{barom1func}), one can make sure that 
\bea
r(3-\la)=-r(\la).
\label{eqforr}
\ena
Hence, if we transform $\la\to 3-\la$ in (\ref{barbaromfunc}) and add both equations,
we come to the relation
\bea
\bar{\bar{\om}}^{(3)}(\la)-\bar{\bar{\om}}^{(3)}(-\la)=
\bar{\bar{\om}}^{(3)}(\la-3)-\bar{\bar{\om}}^{(3)}(3-\la),
\label{relforbarbarom}
\ena
which means that $\bar{\bar{\om}}^{(3)}(\la)-\bar{\bar{\om}}^{(3)}(-\la)$ must
be a periodic function of $\la$ with the period $3$. But from the 
asymptotic behavior (\ref{asympbarbarom}) we see that this function is $0$. 
We thus come to the conclusion 
that the function $\bar{\bar{\om}}^{(3)}(\la)$ must be even
\bea
\bar{\bar{\om}}^{(3)}(\la)=\bar{\bar{\om}}^{(3)}(-\la).
\label{barbaromeven}
\ena
There is a simple case, namely, the point $\la=3/2$, where the left hand side of
(\ref{barbaromfunc}) is $0$ if (\ref{barbaromeven}) is true. Then the right-hand 
side must be $0$ as well. We can explicitly check that, indeed, $r(3/2)=0$. Of 
course, it also follows from (\ref{eqforr}). 

Now, if we take the limit $\la\to 0$ in both sides of equation (\ref{barbaromfunc}), 
we can conclude that the limit of $\bar{\bar{\om}}^{(3)}(\la)$ when $\la\to 0$ and $\la\to\pm3$ 
should exist and
\bea
\bar{\bar{\om}}^{(3)}(3)-\bar{\bar{\om}}^{(3)}(0)=\frac{54-\sqrt{3}\pi-9\log{3}}{4096}.
\label{barbaromat0at3}
\ena
From (\ref{barom3}) we have 
\bea
\bar{\bar{\om}}^{(3)}(\la)=\frac{\om^{(3)}(\la,0,0)}{(\la-\la^{-1})^2}.
\label{barom3om3}
\ena 
Consequently, for $\lambda$ close to zero, we expect 
the following behavior:
\bea
\bar{\bar{\om}}^{(3)}(\la)=\mathcal{O}(\la^{2})\quad {\mbox{when}}\quad  \la\to 0.
\label{barom3at0}
\ena 
It means that the right-hand side of (\ref{barbaromat0at3}) is nothing but 
$\bar{\bar{\om}}^{(3)}(3)$. However, our task 
is more complicated. Actually, we have to calculate the coefficient 
that stands at $\la^2$ on the right-hand side of (\ref{barom3at0}). 

To this end, we proceed to our third step. Here we obtain an integral representation for 
$\bar{\bar{\om}}^{(3)}(\la)$. We will explain some details in Appendix \ref{s:apb}. 
Let us show the final answer:
\bea
&&
\bar{\bar{\om}}^{(3)}(\la)=
-\int\limits_{-i\infty}^{i\infty} 
\frac{d\mu}{6i}\tan\Bigl({\frac{\pi}{3}(\la-\mu)}\Bigr)s(\mu)+t(\la),
\label{barbarom3res}
\ena
where
\begin{multline}
s(\la):= \Biggl(\frac{32\la(4\la^2-13)}{d_1(\la)}+\partial_{\la}\Biggr)
\left\{ \frac{(4\la^2-9)}{16 d_1(\la)} \right. \\
\left. \times \Biggl(\psi\Bigl(\frac12+\frac{\la}3\Bigr)
+\psi\Bigl(\frac12-\frac{\la}3\Bigr)
-\psi\Bigl(\frac56+\frac{\la}3\Bigr)
-\psi\Bigl(\frac56-\frac{\la}3\Bigr)\Biggr) \right\}
\label{s}
\end{multline}
and
\begin{multline}
t(\la) := \frac{\la^2-9}{960(\la^2-1)^2}-
\Bigl(\frac1{240}-\frac45 a\Bigr)\frac{\la^2-4}{(\la^2-1)^2} \\
+ \Bigl(\frac1{960}-\frac4{15}a\Bigr)\Biggl(\psi_1\Bigl(\frac13+\frac{\la}3\Bigr)
+\psi_1\Bigl(\frac13-\frac{\la}3\Bigr)\Biggr) \\
- \Bigl(\frac1{864}-\frac2{9}a\Bigr)\pi^2\Biggl(
\frac1{\sin^2{\Bigl(\frac{\pi}{3}(1+\la)\Bigr)}} +
\frac1{\sin^2{\Bigl(\frac{\pi}{3}(1-\la)\Bigr)}} \Biggr) \\
+ \Biggl(\frac{a+b}3-\frac1{1152}\Biggr)\Biggl(
\cot{\Bigl(\frac{\pi}{3}(1+\la)\Bigr)}+\cot{\Bigl(\frac{\pi}{3}(1-\la)\Bigr)}\Biggr).
\label{t}
\end{multline}
Here we have used the notation
\bea
&& d_1(\la):=(2\la-5)\,(2\la-1)\,(2\la+1)\,(2\la+5)\label{d1}
\ena
and
\bea
&& a=\frac{12-\sqrt{3}\pi-9\log{3}}{2304},\label{ab}\\
&& 
b=\frac{1}{2304}\Bigl(-6+3\sqrt{3}\pi+20\pi^2+27\log{3}-24\,\psi_1\Bigl(\frac13\Bigr)\Bigr).
\nn
\ena

One can verify that 
$\bar{\bar{\om}}^{(3)}(0)=0$, and so, we come to the behavior 
(\ref{barom3at0}) at small $\la$, as expected.

So, we obtain for the limit (\ref{limom3})
the expression 
\begin{multline}
\om^{(3)}(\vec 0)\;=\; \frac{\pi^2}{54\, i}
\int\limits_{-i\infty}^{i\infty} 
d\mu\;\frac{\sin{\frac{\pi\mu}3}}{\cos^3{\frac{\pi\mu}3}}\; s(\mu)
+\frac4{135} \Bigl(\frac1{256} - a\Bigr) \psi_3\bigl(1/3\bigr) - 
\frac{\pi^3}{972 \sqrt{3}} \psi_1\bigl( 1/3\bigr) \\
- \frac{4 (1701 + 20 \sqrt{3} \pi^3 - 40 \pi^4)}{1215} a 
+ \frac{\sqrt{3} \pi^3}{2916}\Bigl(1 + \frac56 \pi^2\Bigr)  
- \frac{\pi^4}{1458} + \frac{11}{960}
\;\simeq\; 0.004579302\ldots 
\label{limom3res}
\end{multline}
Now it is possible to obtain concrete numbers for the elements of the density
matrix $D_{1,2,3}$ defined in (\ref{Delements}). We can use formula (\ref{D3fromGL3})
which allows one to express all matrix elements through six elements
\bea
D_{i_1, i_2, i_3}^{i'_1, i'_2, i'_3} &=& 
D_{1,2,3}^{1,2,3}\; \delta_{i_1}^{i'_1}\,\delta_{i_2}^{i'_2}\,\delta_{i_3}^{i'_3} 
+ D_{2,1,3}^{1,2,3}\; \delta_{i_1}^{i'_2}\,\delta_{i_2}^{i'_1}\,\delta_{i_3}^{i'_3} 
+ D_{1,3,2}^{1,2,3}\; \delta_{i_1}^{i'_1}\,\delta_{i_2}^{i'_3}\,\delta_{i_3}^{i'_2} \nn\\ 
&& + D_{3,2,1}^{1,2,3} \;\delta_{i_1}^{i'_3}\,\delta_{i_2}^{i'_2}\,\delta_{i_3}^{i'_1} 
+ D_{3,1,2}^{1,2,3} \;\delta_{i_1}^{i'_3}\,\delta_{i_2}^{i'_1}\,\delta_{i_3}^{i'_2} 
+ D_{2,3,1}^{1,2,3} \;\delta_{i_1}^{i'_2}\,\delta_{i_2}^{i'_3}\,\delta_{i_3}^{i'_1}.
\label{D3res}
\ena
Then, using the above result (\ref{limom3res}), we can take the homogeneous limit 
in formulas (\ref{resD3})--(\ref{f123}). And so, finally, we 
arrive at the main result of this paper:
\bea
&& D_{1,2,3}^{1,2,3}\;=\;\frac{16}9\om^{(3)}(\vec 0)+\frac1{108}+\frac{\pi^3}{486 \sqrt{3}}+
\frac{\zeta(3)}{27}\;=\; 0.09875519\ldots \nn\\
\nn\\
&& D_{2,1,3}^{1,2,3}\;=\;D_{1,3,2}^{1,2,3}\;=\;
-\frac{8}3\om^{(3)}(\vec 0)+\frac{\pi}{36\sqrt{3}}-\frac{\pi^3}{162\sqrt{3}}+\frac{\log{3}}{12}
-\frac{\zeta(3)}{9}\;=\; -0.0865642\ldots\nn\\
\label{homD3res}\\
&& D_{3,2,1}^{1,2,3}\;=\;
-\frac{8}3\om^{(3)}(\vec 0)+\frac1{36}-\frac{\pi}{72\sqrt{3}}-\frac{\log{3}}{24}
\;=\; -0.0554009\ldots \nn\\
\nn\\
&& D_{3,1,2}^{1,2,3}\;=\;D_{2,3,1}^{1,2,3}\;=\;
4\om^{(3)}(\vec 0)-\frac{\pi}{16\sqrt{3}}+\frac{\pi^3}{108\sqrt{3}}-\frac{3\log{3}}{16}
+\frac{\zeta(3)}{6}\;=\; 0.0650622\ldots\nn
\ena
One can check that, applying the left and right reductions
(\ref{leftrightred}) to (\ref{homD3res}), one reproduces the $n=2$ 
result (\ref{Dhomres}).

As in the case $n=2$, we can test the results by a direct 
numerical diagonalization of the transfer matrix up to the 
length $L=12$ (see Table 1 for $n=2$ and Table 2 for $n=3$ 
cases, respectively):
\begin{table}[h]
	\begin{center}
		\setlength\extrarowheight{7.0pt}
		\begin{tabular}{|c|c|c|c|c|}
			\hline
			&\qquad$D_{1,2,3}^{1,2,3}$ \qquad&\qquad $D_{2,1,3}^{1,2,3}$\qquad &\qquad $D_{3,2,1}^{1,2,3}$\qquad &\qquad $D_{3,1,2}^{1,2,3}$\qquad \\ 
			\hline
			Exact result ($L=\infty $)& 0.09875519 & -0.0865642 & -0.0554009 & 0.0650622 \\ \hline
			\qquad L=9\qquad& 0.103058 & -0.0919341 & -0.0617796 & 0.0713772 \\
			\hline
			\qquad L=12\qquad& 0.10112 & -0.0895192 & -0.0588942 & 0.0685262 \\
			\hline
		\end{tabular}
	\end{center}
	\caption{\label{tablenum} Comparison of numerical and analytic results}
\end{table} 

We see that, unfortunately, the above numerical results cannot provide us 
with sufficient accuracy to really confirm (\ref{limom3res}), since we 
are limited by the small size of the system.  
We actually need some other numerical methods, like DMRG, in order to access $L>12$. We would like to 
return to this problem in the future.

\section{Conclusions}

In this paper, the density matrix of the rational $\sltr$-invariant model 
is explicitly calculated for the operator lengths $n=1,2,3$ 
in both the homogeneous and the inhomogeneous cases (see formulas 
(\ref{Dres}), (\ref{resD3}), (\ref{Dhomres}), (\ref{homD3res})). 
To obtain this result, we had to satisfy the 
reduced qKZ equations (\ref{rqKZ1}), (\ref{rqKZ2}) and all relations 
(i)--(vii) from Section 3. To this end, we had to introduce two transcendental functions 
$\om^{(1)}$ and $\om^{(3)}$. We have studied some of  
their properties. We have established that the first function $\om^{(1)}$ 
is related to the pre-factor $\rho$ of the $R$-matrix given by (\ref{tilderho}),
or to the logarithmic derivative of the $\Gamma$-function $\psi$. In this sense it looks rather 
similar to the function $\om$ defined in \cite{BJMT040506} in the $\slt$-case. Hence, it was not 
a big surprise that we faced a particular value of Riemann's
zeta-function which appeared when we took the homogeneous limit. But the other function $\om^{(3)}$ 
is certainly more non-trivial. In the homogeneous limit it is related to an integral
of Fourier type, where the integrand itself is expressed in terms of the $\psi$-function and its 
derivatives. In this sense we confirm the result by Martin and Smirnov \cite{MarSmi} that one 
cannot expect the factorization to single integrals of the elementary functions. Actually, the 
paper \cite{MarSmi} is devoted to the consideration of the classical integrable model with a 
non-hyperelliptic spectral curve associated with the $\sltr$-symmetry. Of course, it is 
not directly related to the result of our paper. As we believe, we should consider more 
general correlation functions corresponding to any representation with arbitrary weights 
associated with the quantum space. One might expect that in some special limit, when such 
weights tend to infinity, we could access the classical limit. Still, we may expect that 
we can extract certain information about an overall structure, such 
as the number of non-trivial transcendental functions which was found 
to be six in \cite{MarSmi}. 
And so, we think that we will also face more non-trivial 
transcendental functions when we will consider operators 
of length larger than $3$.

We believe that  we have to generalize the factorization structure that 
we faced in the $\slt$-case in such a sense that in order to 
describe general correlation functions of the $\sltr$-invariant model we 
have to introduce more transcendental functions which cannot be related 
to one-dimensional integrals of elementary functions with the coefficients 
of algebraic nature. Still, we think the result of this paper is rather 
encouraging that this program can be realized. This is our future project.

\appendix

\section{Reduced qKZ equations: heuristic derivation} \label{s:apa}

Here we describe the method of obtaining the reduced qKZ equations using pictures. 
Essentially, we follow the method used by Aufgebauer and Kl{\"u}mper for the derivation 
of the discrete rqKZ relation in the $\slt$-case \cite{AufKlum}.

We denote by black lines the fundamental and by blue lines the anti-fundamental representations. 
A cross denotes the ``charge conjugation'' operator (see (\ref{C})). Introduce the graphical 
notation for the main $R$-matrix properties: initial condition and crossing relations (see 
(\ref{RbarRspecial})--(\ref{cross2}))
\begin{figure}
\includegraphics[scale=1.0]{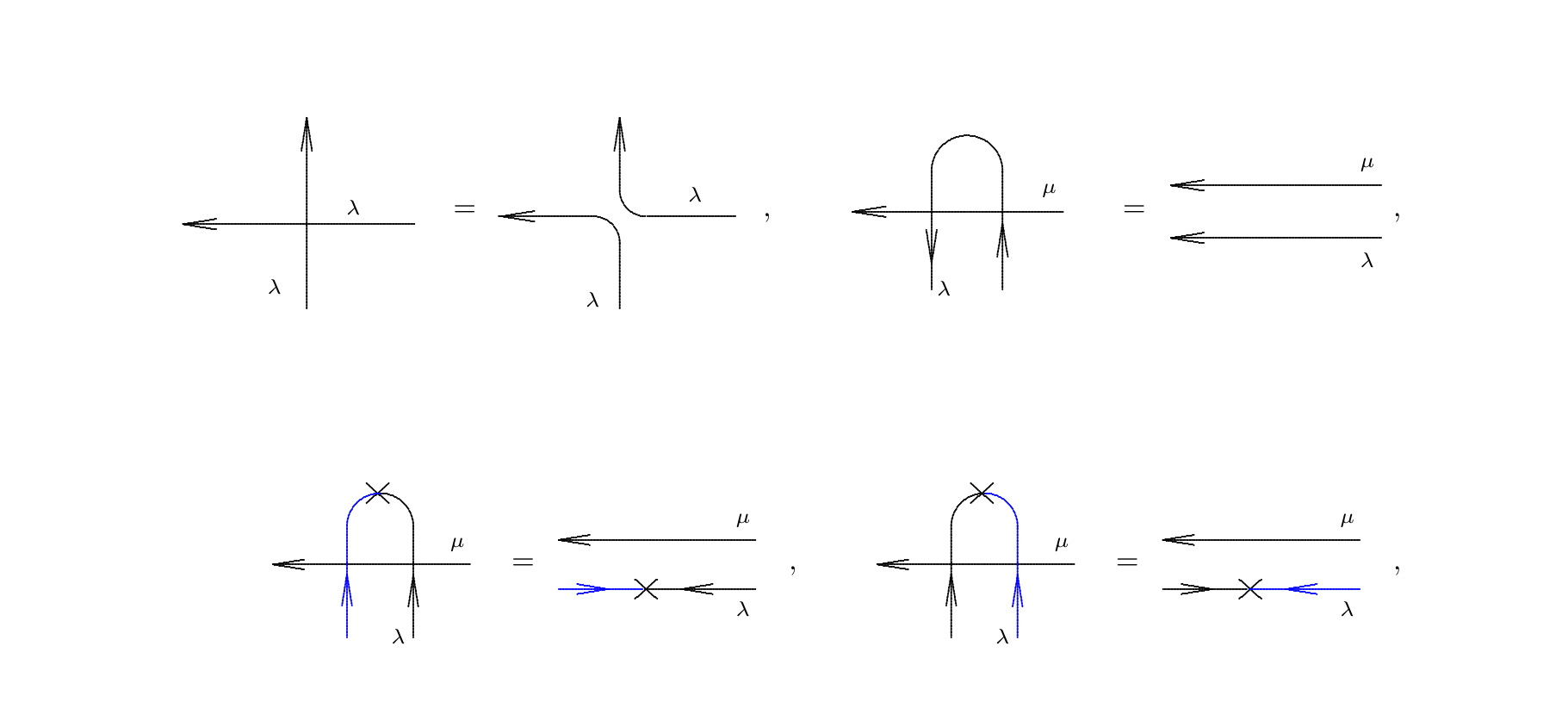}
\caption{R-matrix properties}
\label{rmatprop}
\end{figure}
 (all the crossing relations remain true if the horizontal line is blue (anti-fundamental)) 
 and the conjugation identities (\ref{barR}) are depicted in Fig.~\ref{rmatprop} and Fig.~2.
\begin{figure}
\label{crossing1}
\includegraphics[scale=1.0]{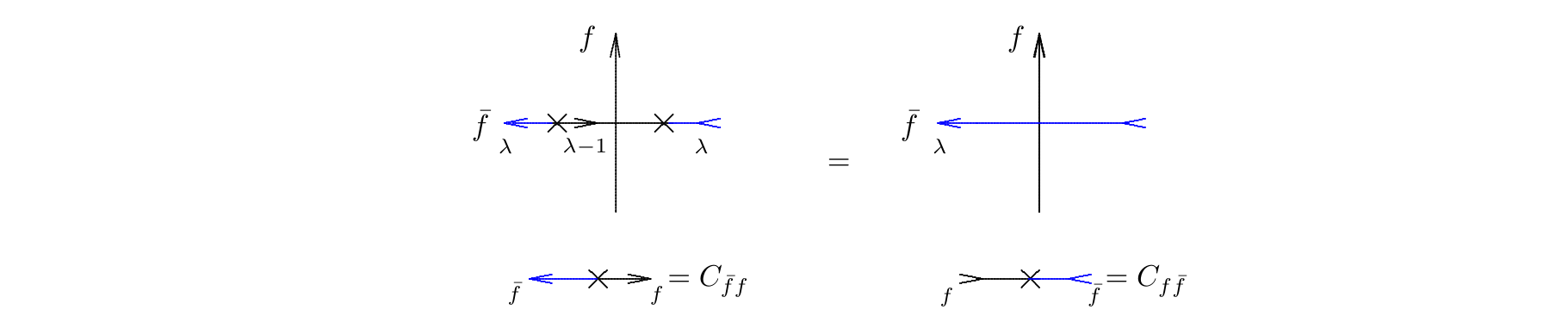}
\end{figure}
\begin{figure}[htbp]
\includegraphics[scale=1.0]{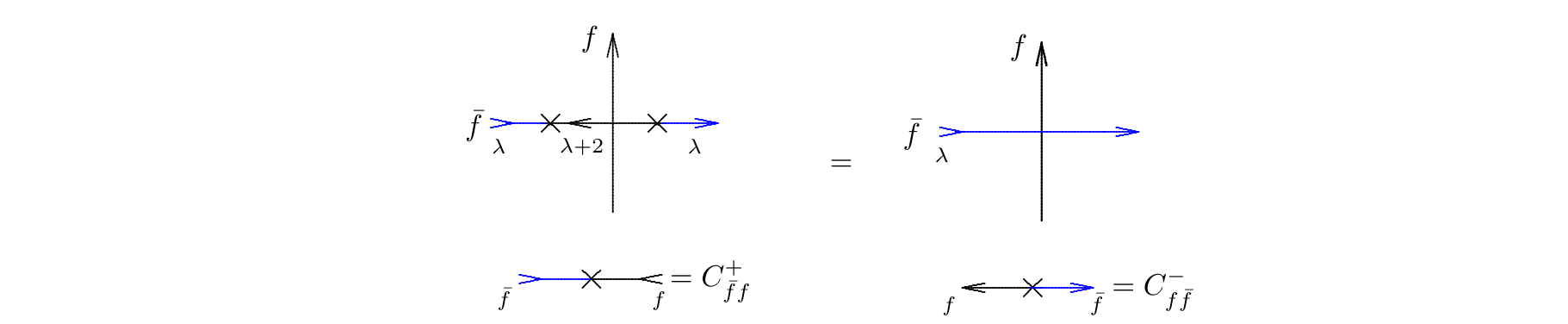}
\caption{Crossing relations}
\label{crossing2}
\end{figure}%
Let us start with a more general case of the construction 
which is used for the introduction of temperature (see for example 
the book \cite{EssFraGoeKluKor}). We consider a lattice which contains 
an additional direction, sometimes called the Matsubara direction as was 
discussed in paper \cite{HGSIII}. Periodic boundary conditions in both, 
vertical and horizontal, directions are implied. So, we take $2N$ lattice 
sites in the Matsubara direction, where the corresponding 
horizontal lines are taken in staggered order as it is shown in  
Fig.~\ref{staggering}, \cite{MSuzuki,Trotter,Klumper}.

Obviously, we can depict the $n$-site density matrix 
$D^{i_1'\dots i_n'}_{i_1\dots i_n}(\lambda_1,\dots,\lambda_n)$ 
as shown on Fig.~\ref{staggering}. 
\begin{figure}[htbp]
\includegraphics[scale=1.0]{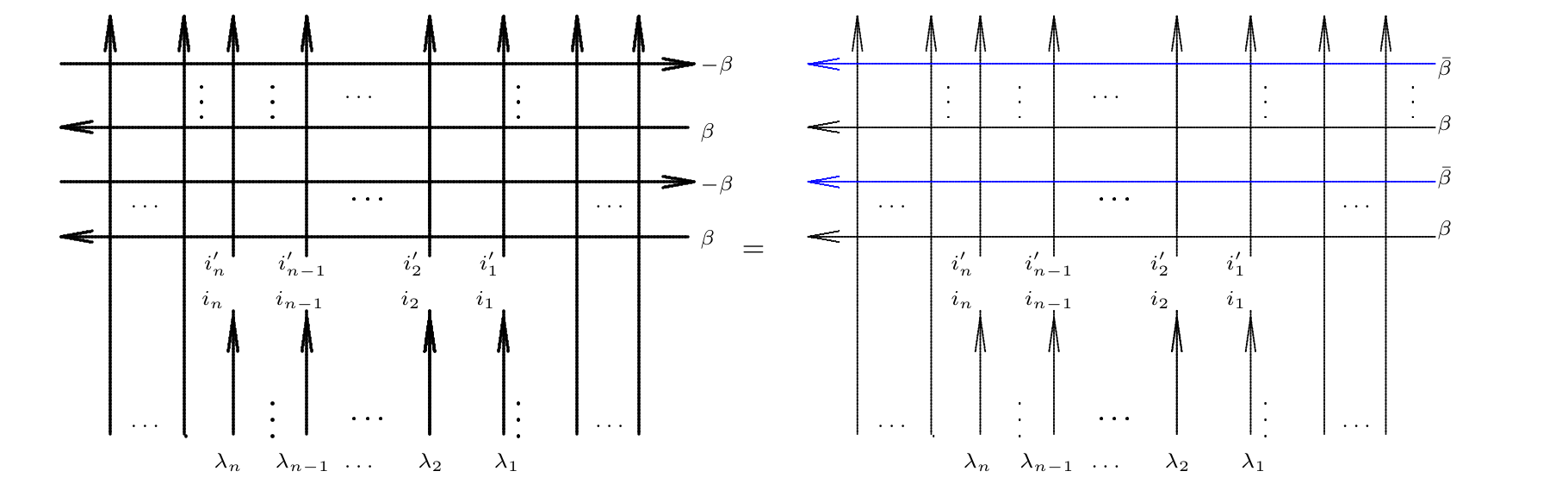}
\caption{Introducing the anti-fundamental representation}
\label{staggering}
\end{figure}
Due to periodic boundary conditions in both, vertical and horizontal, directions, 
the lines form closed loops, except the lines with the cut. The density matrix 
acts onto $n$-sites quantum operators that graphically can be inserted into the 
cut with edges $i_1,\dots, i_n, i_1',\dots,i_n'$. All transfer 
matrices commute, so we can change the line order every moment. The horizontal 
(the auxiliary spaces) lines correspond to usual transfer matrices $T(\lambda)$, 
and the vertical (the quantum spaces) ones correspond to the quantum 
transfer matrices. Let us set $\beta=1/(2TN)$, where $T$ is the temperature. We 
intend to work with the zero temperature case, so we must be careful with the limit 
$N\rightarrow\infty$. The normalization to the highest eigenvalue is implied for all 
the transfer matrices $T(\lambda)$. We will come to this point later.
 
The $N$ lines going from the left to the right can be rewritten using the 
crossing relations (\ref{barR}) and associated anti-fundamental representations 
in such a way that all $2N$ horizontal lines go from the right to the left, where 
$N$ lines correspond to the fundamental representation, and the other $N$ lines 
correspond to the anti-fundamental representation, as shown in  
Fig.~\ref{staggering}.

In the spirit of the paper \cite{HGSIII}, we can consider a more general case with 
two sets of arbitrary spectral parameters $\{\beta_j\}, \{\bar{\beta}_j\}, j=1,\cdots N$, 
associated with $N$ ``fundamental lines'' and $N$ ``anti-fundamental lines'', respectively. 

Then we can deduce two kinds of relations:
one set for the case when $\la_1=\beta_j, j=1,\cdots N$ and another set
for $\la_1={\bar\beta}_j, j=1,\cdots N$. Since the transfer matrices corresponding
to horizontal lines commute, we can take any fixed number $j$ for our derivation
without any loss of generality. After further manipulations shown in 
Fig.~\ref{qkzderiv1}--Fig.~\ref{qkzderiv2}, we will come to a couple 
of two relations which connect two objects earlier shown in Fig.~\ref{density01}:
\bea
 D^{(0)}
(\la_1,\ldots,\la_n|\beta_1,\ldots,\beta_N;\bar{\beta}_1,\ldots,
\bar{\beta}_N) \qquad {\mathrm{and}} \qquad 
 D^{(1)} 
(\la_1,\ldots,\la_n|\beta_1,\ldots,\beta_N;\bar{\beta}_1,\ldots,
\bar{\beta}_N). \nn
\ena
The first object $D^{(0)}$ corresponds to the situation,
where the fundamental representation is assigned to all 
vertical lines. The second object $D^{(1)}$ corresponds to
the situation with one ``anti-fundamental'' vertical line 
taken for the space $1$, while all other vertical lines 
are ``fundamental''.
\begin{figure}
\includegraphics[scale=1.0]{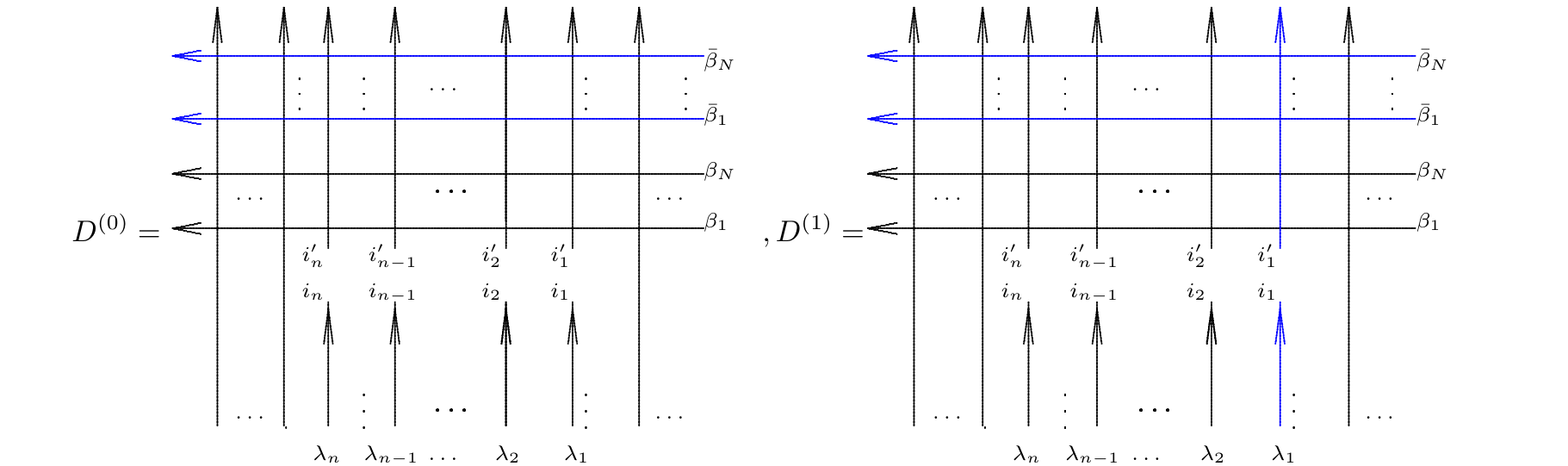}
\caption{Two density matrices}
\label{density01}
\end{figure}

Now we are going to prove that these density matrices satisfy 
the equations shown in Fig.~\ref{qkz1} and Fig.~\ref{qkz2}. 
The relations can easily be established using pictures.
A single action of the $A^{(1)}$-operator (see definition 
(\ref{A1})) onto the matrix $D^{(0)}$ can be shown by 
Fig.~\ref{qkzderiv1}.
\begin{figure}
\includegraphics[scale=1.0]{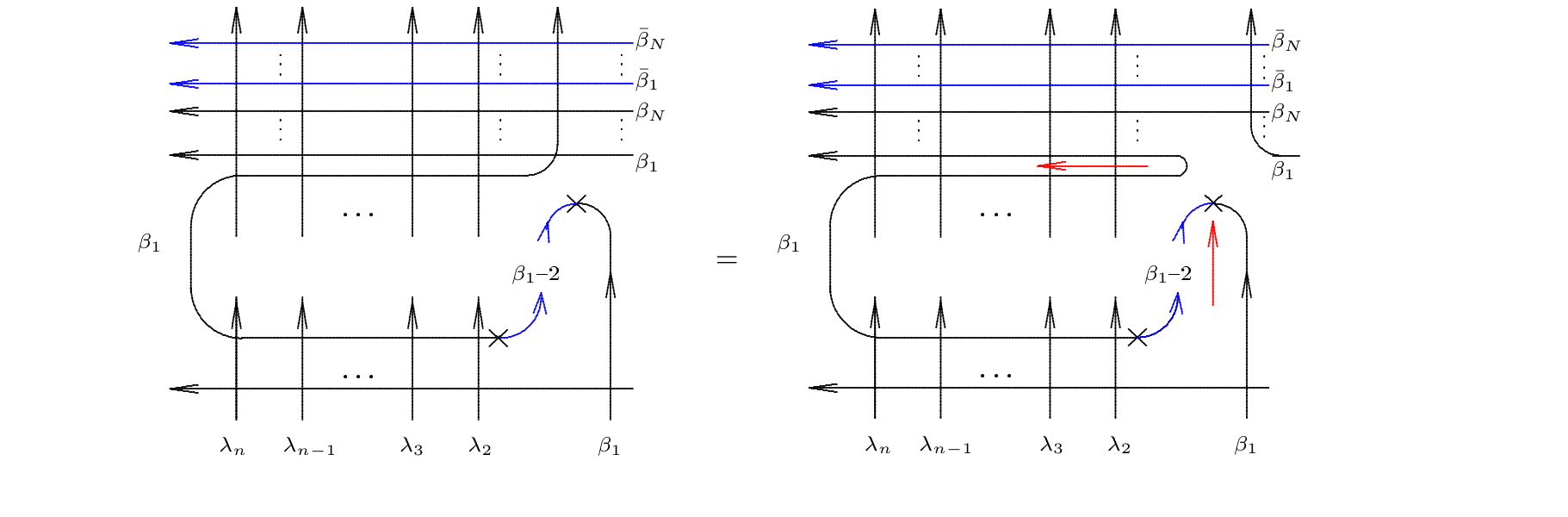}
\caption{Relation for $D^{(0)}$}
\label{qkzderiv1}
\end{figure}
Here we use initial condition $R(0)=P$. In order not to overload pictures, 
let us omit all the quantum spaces here except those containing the cut, but 
keep in mind their existence. In the last expression we can make a few 
transformations. Firstly, we move the left ``curve'' through all the 
vertical lines (see red arrow on the picture). We can do it just using 
the unitarity relation for the $R$-matrices. Next, using the 
crossing relations (\ref{cross1}, (\ref{cross2}), we move the remaining 
``curve'' through all the horizontal lines in vertical direction (see 
the red arrow), then such a loop comes back in the bottom due to the 
cyclicity, and also one additional vertical line appears.\\
\begin{figure}
\includegraphics[scale=1.0]{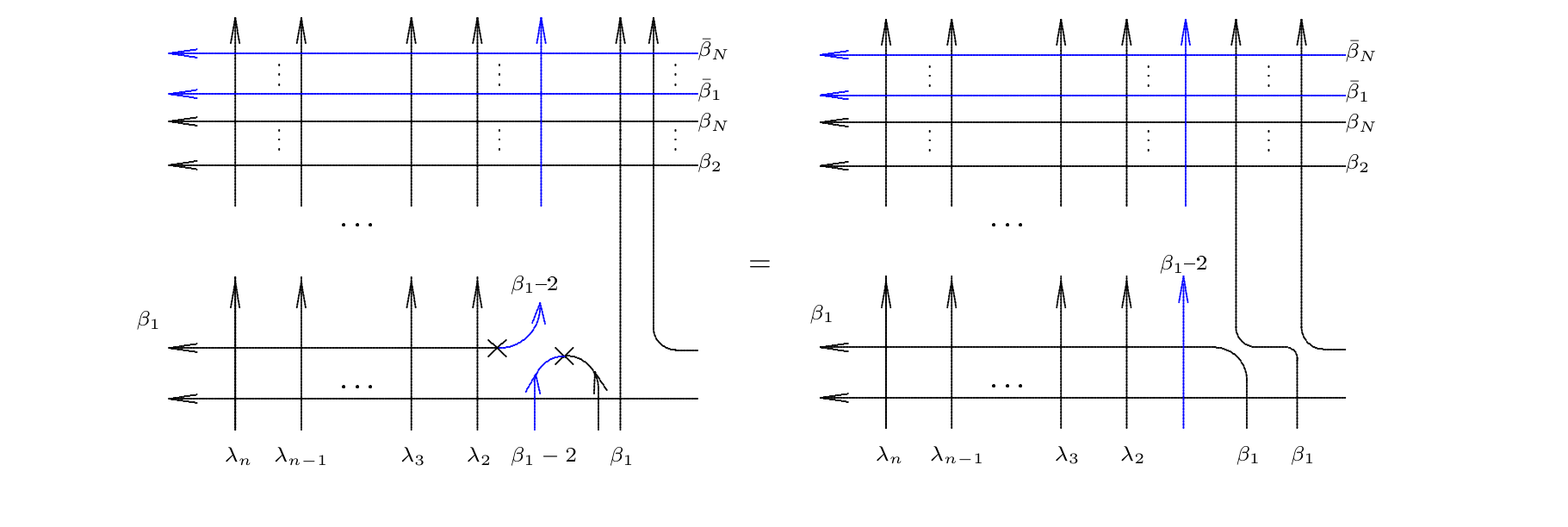}
\caption{Relation for $D^{(1)}$}
\label{qkzderiv2}
\end{figure}
Now let us use again the initial condition Fig.~\ref{rmatprop}  in order to obtain 
the final expression. At the last step we use the initial condition two times more 
for the last three lines, and the density matrix restores automatically at the 
right-hand side of the relation. Formally, there is a problem concerning two additional 
quantum spaces that appear in the final expression. But actually, this is not a 
problem because at the end we send the number of the quantum spaces to infinity 
(thermodynamic limit), and so, the few additional lines do not change anything 
if they are properly normalized. In our case we have to normalize two additional 
vertical lines. It causes the appearance of the factor 
$\Lambda(\beta_1)\bar\Lambda(\beta_1-2)$. Here $\Lambda(\la)$ and $\bar\Lambda(\la)$ 
are maximal eigenvalues of the quantum transfer matrices corresponding to fundamental 
and anti-fundamental representations, respectively. Fortunately, the above factor 
is just equal to $1$. Therefore, we get the correct 
normalization automatically.
  
Finally, we obtain the relation depicted in Fig.~\ref{qkz1}.
\begin{figure}
\includegraphics[scale=1.1]{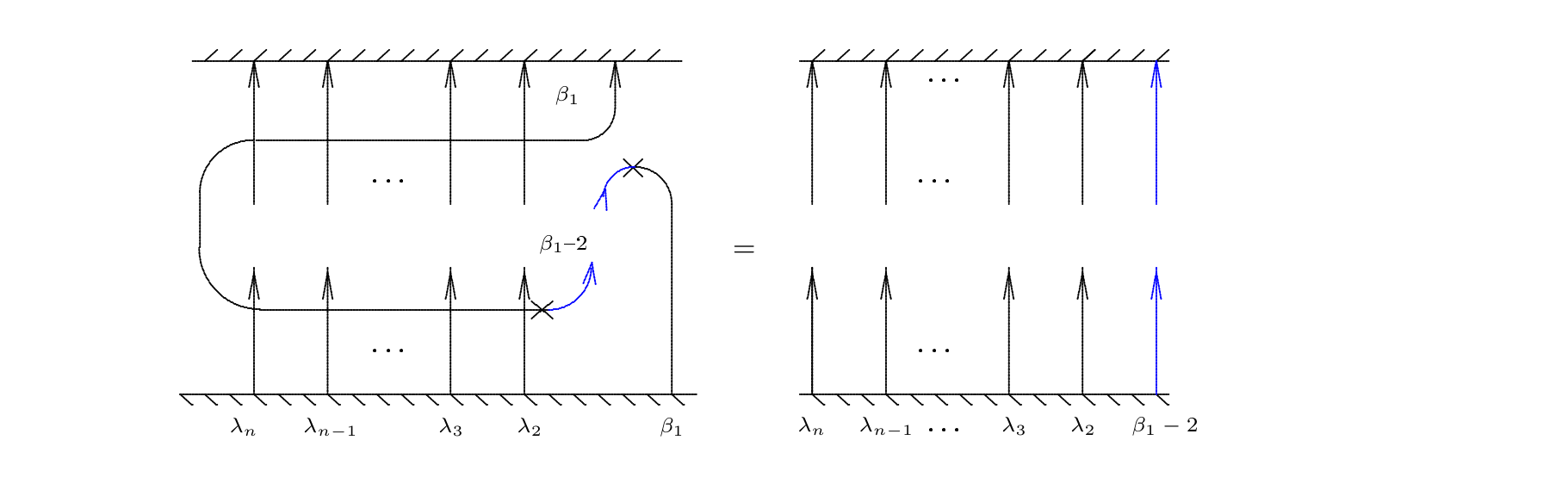}
\caption{First difference equation}
\label{qkz1}
\end{figure}

As we already pointed out, the set of quantum spaces without cut are implied, 
but they are not shown in the picture. Now we denote 
an infinite set of auxiliary spaces just by the grounding. 

 We wish to stress that on the right-hand side of the relation 
	we get the matrix\\ $D^{(1)}(\la_1,\ldots,\la_n|\beta_1,\ldots,\beta_N;\bar{\beta}_1,\ldots,
	\bar{\beta}_N)$ with the proper shift of spectral parameter and with one
	anti-fundamental line corresponding to the first quantum space.
	Meanwhile on the left hand side we have the matrix
	$D^{(0)}(\la_1,\ldots,\la_n|\beta_1,\ldots,\beta_N;\bar{\beta}_1,\ldots,
	\bar{\beta}_N)$ we started with.

	
	 In a similar way we can obtain the second equation using the 
	operator (\ref{A2}) and $D^{(1)}$ (see Fig.~\ref{qkz2}).
\begin{figure}
\includegraphics[scale=1.1]{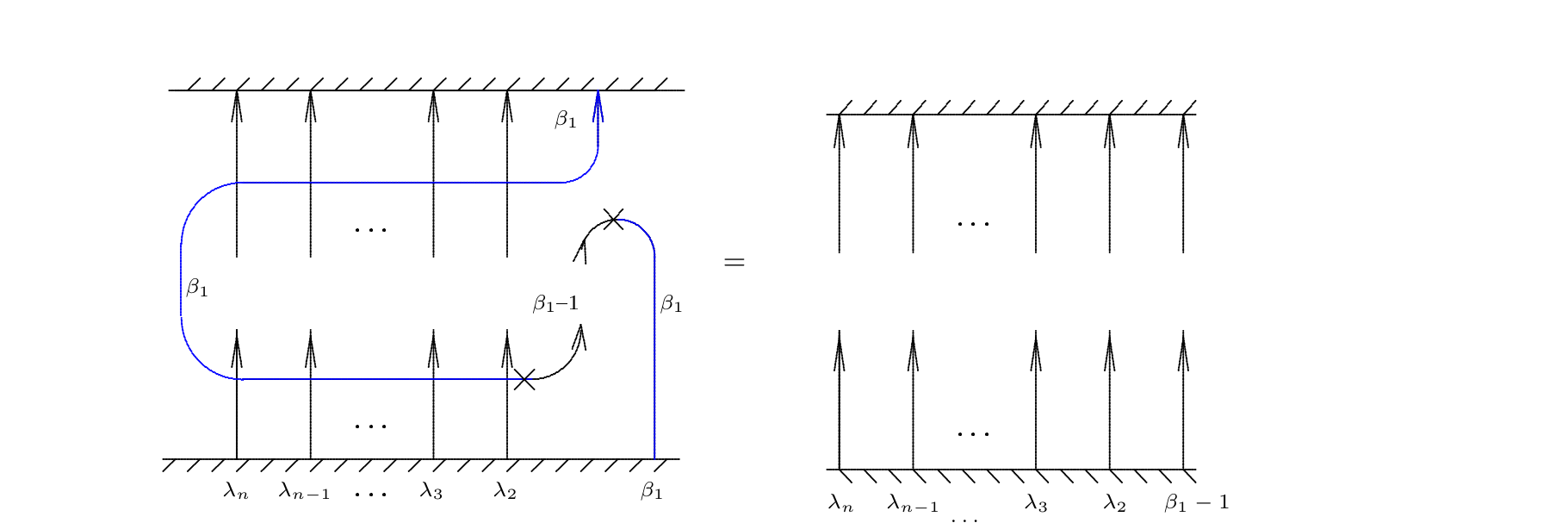}
\caption{Second difference equation }
\label{qkz2}
\end{figure}

 Then we have the density matrix $D^{(1)}(\la_1,\ldots,\la_n|\beta_1,\ldots,\beta_N;\bar{\beta}_1,\ldots,
	\bar{\beta}_N)$ on the left-hand side and  $D^{(0)}(\la_1,\ldots,\la_n|\beta_1,\ldots,\beta_N;\bar{\beta}_1,\ldots,
	\bar{\beta}_N)$ on the right-hand side of our 
	relation.

Finally, we come to the following two equations:
\begin{multline}
D^{(0)} 
(\beta_j,\la_2,\ldots,\la_n|\beta_1,\ldots,\beta_N;\bar{\beta}_1,\ldots,
\bar{\beta}_N)
\Bigl(A^{(1)}_{1,\bar{1}|2,\ldots,n}(\beta_j|\la_2,\ldots,\la_n)
\bigl(X_{\bar{1},2,\ldots,n}\bigr)\Bigr) \\
= D^{(1)} 
(\beta_j-2,\la_2,\ldots,\la_n|\beta_1,\ldots,\beta_N;\bar{\beta}_1,\ldots,
\bar{\beta}_N)\bigl(X_{\bar{1},2,\ldots,n}\bigr),
\label{rqKZ1withbeta}
\end{multline}
\begin{multline}
D^{(1)} 
({\bar\beta}_j,\la_2,\ldots,\la_n|\beta_1,\ldots,\beta_N;\bar{\beta}_1,\ldots,
\bar{\beta}_N)
\Bigl(A^{(2)}_{\bar{1},1|2,\ldots,n}({\bar\beta}_j|\la_2,\ldots,\la_n)
\bigl(X_{1,\ldots,n}\bigr)\Bigr) \\
= D^{(0)} 
({\bar\beta}_j-1,\la_2,\ldots,\la_n|\beta_1,\ldots,\beta_N;\bar{\beta}_1,\ldots,
\bar{\beta}_N)\bigl(X_{1,\ldots,n}\bigr).
\label{rqKZ2withbeta}
\end{multline}

Let us comment on the above relations (\ref{rqKZ1withbeta}), (\ref{rqKZ2withbeta}) 
 (see also Fig.~\ref{qkz1} and Fig.~\ref{qkz2}). 
Actually, we would not call them rqKZ equations since they are not really difference relations. 
The expressions standing on the left- and on
the right-hand side are essentially different because the first argument of $D^{(0)}$ and $D^{(1)}$, 
$\beta_j$ or $\bar{\beta}_j$ respectively, is shifted on the right-hand side 
of (\ref{rqKZ1withbeta}), (\ref{rqKZ2withbeta}), but there is also dependence on the non-shifted parameters 
within both sets $\{\beta_j\},\{\bar{\beta}_j\}$, while on the left-hand side of 
(\ref{rqKZ1withbeta}), (\ref{rqKZ2withbeta}) the first argument is not shifted. It means that 
relations (\ref{rqKZ1withbeta}), (\ref{rqKZ2withbeta}) are not closed.

Let us see that in the zero temperature limit we will come to real difference relations.
So, we set again $\beta_k\to\frac1{2TN}$ and ${\bar\beta}_k\to 1-\frac1{2TN}$
for $k=1,\cdots,N$ and then take the so-called 
Trotter limit $N\to\infty$ \cite{Trotter}.  Finally, we have to take the limit $T\to 0$. 

Let us make one important remark here. We could start with the situation 
 	where the whole number of quantum spaces 
 	in horizontal direction is finite, say, $L$. We are interested in taking the limit when both $L$ and $N$ tend to infinity. 
 	In paper \cite{HGSIII} the limit $L\to\infty $ was taken first. Let us assume that both limits commute. So, if we first take the limit $L\to\infty$ keeping the Trotter number $N$ finite, we can insert the projector to the ground state somewhere at 
 	the right or at the left infinity in the vertical direction. We see that in this case all vertical lines except
 	$n$ lines, where the cut is taken disappear if they are properly normalized 
 	by the maximal eigenvalue of the corresponding quantum transfer matrices as we discussed above. On the other hand, 
 	if we take the limit $N\to\infty$, $ T\to 0$ first keeping $2T N=1/\la_1$ and $L$ finite (with
 	some arbitrary parameter $\la_1$), we can insert the projector to the vacuum  somewhere at infinity 
 	in the horizontal direction. Then the horizontal lines will disappear. Again we have
 	to normalize them by putting the maximal eigenvalue into the denominator. 
 	Because of the duality of both pictures we can realize that both ways
 	of normalization are compatible with each other. 
 We have checked numerically that the above scheme actually works well (see Tables 1 and 2).

So, we come to the vacuum expectation value (\ref{cor}):
\bea
&& D^{(0)} 
(\beta_j,\la_2,\ldots,\la_n|\beta_1,\ldots,\beta_N;\bar{\beta}_1,\ldots,
\bar{\beta}_N)\bigl(X_{\bar 1,2,\ldots,n}\bigr)\to 
D 
(\la_1,\ldots,\la_n)\bigl(X_{\bar 1,2,\ldots,n}\bigr),\nn\\
&& D^{(1)} 
(\beta_j,\la_2,\ldots,\la_n|\beta_1,\ldots,\beta_N;\bar{\beta}_1,\ldots,
\bar{\beta}_N)\bigl(X_{1,2,\ldots,n}\bigr)\to 
D^{(1)} 
(\la_1,\ldots,\la_n)\bigl(X_{1,2,\ldots,n}\bigr)
\nn
\ena
with the original generalized density matrix $D$ defined by (\ref{corinhom})
and $D^{(1)}$ can be defined in a similar way  by taking the 
anti-fundamental representation for the first line.  

 Now, as was explained in Section 4, we can combine  equations 
(\ref{rqKZ1withbeta}), (\ref{rqKZ2withbeta}) and get finally the closed equation with the same matrix $D$ at both sides (see \ref{rqkz}).
The corresponding diagram is shown in Fig.~\ref{rqkzfinal} .
\begin{figure}[H]
\includegraphics[scale=1.1]{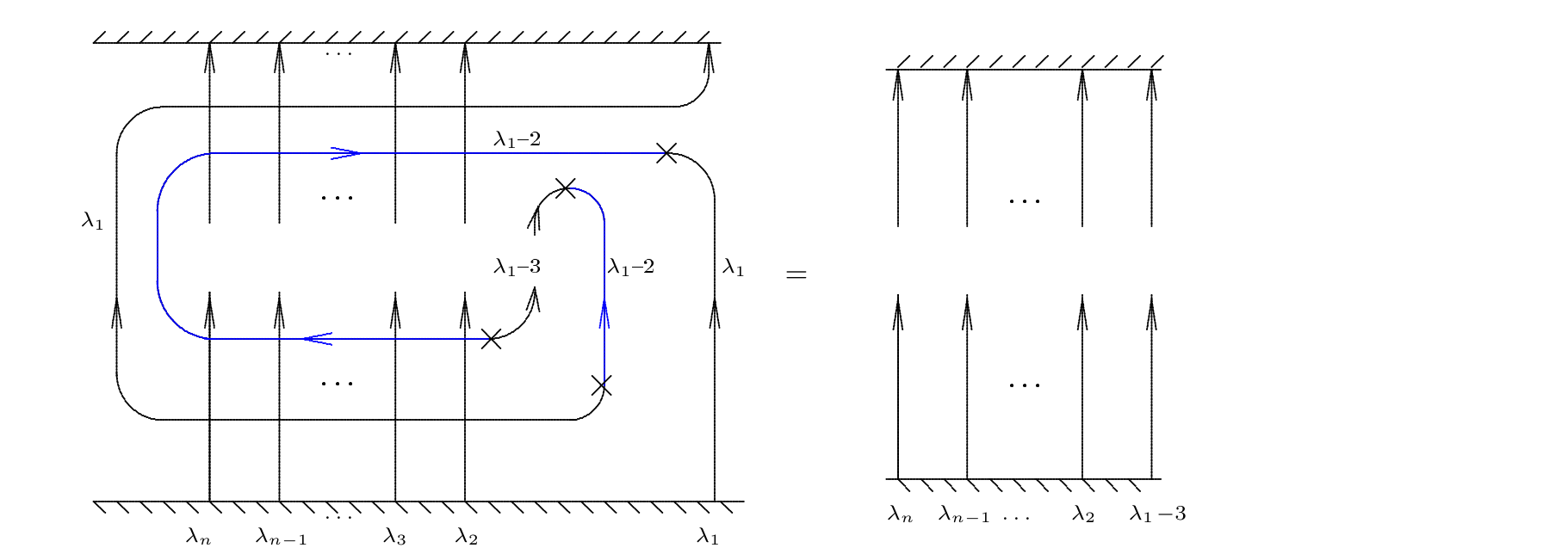}
\caption{rqKZ equation}
\label{rqkzfinal}
\end{figure}

\section{Some details on formula (\ref{barbarom3res})} \label{s:apb}

In this Appendix we explain some details of our derivation of formula (\ref{barbarom3res}) 
from Section 4.4. Let us start with the functional relation (\ref{barbaromfunc}). Then we can substitute 
formula (\ref{barom1c}) for the function $\bar{\om}^{(1)}$ into the right-hand side 
of (\ref{barbaromfunc}) using formula (\ref{r}) for the function $r(\la)$. Applying 
partial fraction decomposition, we can write it in the following form:
\begin{multline}
r(\la)=\sum\limits_{j\ge0}s_j(\la) \\
 -\frac{1}{256}\Bigl(\frac{3/5}{(\la-4)^2}+\frac{3}{(\la-2)^2}-\frac{3}{(\la-1)^2}-
\frac{3/5}{(\la+1)^2}-\frac1{\la-4}+\frac1{\la-2}+\frac1{\la-1}-
\frac1{\la+1}\Bigr)\bar{\om}^{(1)}(0) \\
 -\frac{1}{384}\Bigl(\frac{3/5}{(\la-4)^2}-\frac{1}{(\la-2)^2}-\frac{10}{(\la-1)^2}+
\frac{2/5}{(\la+1)^2}-\frac{19/15}{\la-4}+\frac{13}{\la-2}-\frac{34/3}{\la-1}-
\frac{2/5}{\la+1}\Bigr),
\label{fracr}
\end{multline}
where
\begin{multline}
s_j(\la)=\frac{b_{2,0,j}}{(\la-3j)^2}-\frac{b_{2,0,j+1}}{(\la+3j)^2}
+\frac{b_{2,1,j}}{(\la-1-3j)^2}-\frac{b_{2,1,j+1}}{(\la+1+3j)^2} \\
+\frac{b_{1,0,j}}{\la-3j}+\frac{b_{1,0,j+1}}{\la+3j}
+\frac{b_{1,1,j}}{\la-1-3j}+\frac{b_{1,1,j+1}}{\la+1+3j}\label{sj} \\
+\frac{a_{2,4,j}}{(\la-4)^2}+\frac{a_{2,2,j}}{(\la-2)^2}+\frac{a_{2,1,j}}{(\la-2)^2}
+\frac{a_{2,1,j}}{(\la-1)^2}+\frac{a_{2,-1,j}}{(\la+1)^2}
+\frac{a_{2,-1,j}}{(\la-1)^2}+\frac{a_{2,-1,j}}{(\la-1)^2},
\end{multline}
and
\bea
&&
b_{2,0,j}=\frac1{64}\Bigl(-\frac{2/5}{3j-4}+\frac{2/5}{3j+1}-\frac{1}{3j-2}+\frac{1}{3j-1}\Bigr)\label{ba},\nn\\
&&
b_{2,1,j}=\frac1{64}\Bigl(\frac{2/5}{3j-3}-\frac{2/5}{3j+2}+\frac{1}{3j-1}-\frac{1}{3j}\Bigr),\nn\\
&&
b_{1,0,j}=\frac1{64}\Bigl(\frac{4/5}{(3j-4)^2}-\frac{4/5}{(3j+1)^2}+\frac{2}{(3j-2)^2}-
\frac{2}{(3j-1)^2}
+\frac{1}{3j-4}+\frac{1}{3j+1}-\frac{1}{3j-2}-\frac{1}{3j-1}\Bigr),\nn\\
&&
b_{1,1,j}=\frac1{64}\Bigl(-\frac{4/5}{(3j-3)^2}+\frac{4/5}{(3j+2)^2}-\frac{2}{(3j-1)^2}+\frac{2}{(3j)^2}
-\frac{1}{3j-3}-\frac{1}{3j+2}+\frac{1}{3j-1}+\frac{1}{3j}\Bigr)\nn\\
&&
a_{2,4,j}=\frac{1}{160} \Bigl(-\frac1{3j-4} -\frac{1}{3 j+4} + \frac1{ 3 j+5} + \frac1{3j-3}\Bigr),\nn\\
&&
a_{2,2,j}=\frac{1}{64} \Bigl(-\frac1{3j-2} -\frac{1}{3 j+2} + \frac1{ 3 j+3} + \frac1{3j-1}\Bigr),\nn\\
&&
a_{2,1,j}=\frac{1}{64} \Bigl(-\frac1{3j-1} -\frac{1}{3 j+1} - \frac1{ 3 j+2} -\frac1{3j}\Bigr),\nn\\
&&
a_{2,-1,j}=\frac{1}{160} \Bigl(\frac1{3j+1} +\frac{1}{3 j-1} - \frac1{ 3 j+2} -\frac1{3j}\Bigr),\nn\\
&&
a_{1,4,j}=\frac{1}{80} 
\Bigl(-\frac{1}{(3j-4)^2} + \frac{1}{(3 j+4)^2} - \frac1{ (3 j+5)^2} + \frac1{(3j-3)^2}\Bigr)\nn\\
&& \hskip3cm
+\frac{1}{64}\Bigl(-\frac1{3j-4} -\frac{1}{3 j+4} + \frac1{ 3 j+5} + \frac1{3j-3}\Bigr),\nn\\
&&
a_{1,2,j}=\frac{1}{64} 
\Bigl(-\frac{2}{(3j-2)^2} +\frac{2}{(3 j+2)^2} - \frac2{ (3 j+3)^2} + \frac2{(3j-1)^2}\nn\\
&& \hskip3cm 
+\frac1{3j-2} +\frac{1}{3 j+2} - \frac1{ 3 j+3} - \frac1{3j-1}\Bigr),\nn\\
&&
a_{1,1,j}=\frac{1}{64} 
\Bigl(\frac{2}{(3j-1)^2} -\frac{2}{(3 j+1)^2} +\frac2{ (3 j+2)^2} -\frac2{(3j)^2}
+\frac1{3j-1} +\frac{1}{3 j+1} - \frac1{ 3 j+2} - \frac1{3j}\Bigr),\nn\\
&&
a_{1,-1,j}=\frac{1}{80} 
\Bigl(\frac{1}{(3j+1)^2} -\frac{1}{(3 j-1)^2} +\frac1{ (3 j)^2} - \frac1{(3j+2)^2}\Bigr)\nn\\
&& \hskip3cm 
+\frac{1}{64}\Bigl(-\frac1{3j+1} -\frac{1}{3 j-1} -\frac1{ 3 j} + \frac1{3j+2}\Bigr).
\ena
It is easy to check that
$$
b_{2,0,0}=b_{2,0,1}=b_{1,0,0}=b_{1,0,1}=0.
$$
Besides, it is implied that in the summation (\ref{fracr})
\bea
b_{2,1,0}=b_{2,1,1}=b_{1,0,0}=b_{1,0,1}=b_{1,1,0}=b_{1,1,1}=0.
\label{boundary1}
\ena 
From formulas (\ref{barbaromfunc}), (\ref{fracr}) we can deduce 
that the solution for $\bar{\bar{\om}}^{(3)}$ has the following form:
\bea
\bar{\bar{\om}}^{(3)}(\la)&=&
\;\;\sum\limits_{j\ge 2}c_{2,0,j}\Bigl(\frac1{(\la-3j)^2}+\frac1{(\la+3j)^2}\Bigr)
+c_{1,0,j}\Bigl(\frac1{\la-3j}-\frac1{\la+3j}\Bigr) \label{barbarom3ansatz}\\
&& + \sum\limits_{j\ge 0}c_{2,1,j}\Bigl(\frac1{(\la-1-3j)^2}+\frac1{(\la+1+3j)^2}\Bigr)
+ c_{1,1,j}\Bigl(\frac1{\la-1-3j}-\frac1{\la+1+3j}\Bigr),\nn
\ena
where
\bea
& c_{2,0,j}=-\sum\limits_{k=2}^{j}b_{2,0,k},\qquad 
c_{1,0,j}=-\sum\limits_{k=2}^{j}b_{1,0,k},\label{c}\\
& c_{2,1,j}=-\sum\limits_{k=0}^{j}b_{2,1,k},\qquad 
c_{1,1,j}=-\sum\limits_{k=0}^{j}b_{1,1,k},\nn
\ena
but we should correct the boundary values (\ref{boundary1}). We take 
\bea
&& b_{2,1,j}=\begin{cases}
	a, & j=0\\

		d+\frac{a}5, & j=1\\
\frac1{64}\Bigl(-\frac{4/5}{(3j-3)^2}+\frac{4/5}{(3j+2)^2}-\frac{2}{(3j-1)^2}+
\frac{2}{(3j)^2}
-\frac{1}{3j-3}-\frac{1}{3j+2}+\frac{1}{3j-1}+\frac{1}{3j}\Bigr),
 & j\ge2			
\end{cases}
\nn\\
\label{boundary3}\\
&& 
b_{1,1,j}=\begin{cases}
b, & j=0\\

c-\frac{2}5b, & j=1\\
\frac1{64}\Bigl(\frac{2/5}{3j-3}-
\frac{2/5}{3j+2}+\frac{1}{3j-1}-\frac{1}{3j}\Bigr), & j\ge2			
\end{cases}.
\nn
\ena
In Section 4.4 we defined the values of $a$ and $b$ by the formulas (\ref{ab}).
Also we should define
\bea
&& c=\frac{2644-225\sqrt{3}\pi-2025\log{3}}{28800}\label{cd},\\
&& d=-\frac{77}{9600}.\nn
\ena

One can check that with this definition one has the coinciding residues 
of the left- and right-hand sides of the functional relation
(\ref{barbaromfunc}) at all values $\la=3j,\;\pm1+3j$,  where  
$j=0,\pm1,\pm2,\cdots$. Explicitly one has from (\ref{barbarom3ansatz}) 
$$
\bar{\bar{\om}}^{(3)}(\la)=\bar{\bar{\om}}^{(3)}(-\la).
$$
Also one can see that it falls down at least as 
$1/\la^{2}$ when $\la\to \infty$. So, using the Liouville theorem we come to the 
conclusion that the solution (\ref{barbarom3ansatz}) for  $\bar{\bar{\om}}^{(3)}(\la)$ 
is unique. 

In order to come to the form (\ref{barbarom3res}), we can 
separate the sum on the right-hand side of (\ref{barbarom3ansatz}) 
in two parts 
\bea
&&\bar{\bar{\om}}^{(3)}(\la)=I_0(\la)+I_1(\la),
\label{barbarom3res1}
\ena
where the summation in the first term $I_0$ goes from 2 to $\infty$ and
the summation in the second term $I_1$ goes from 0 to 1. 

So, we should take the corresponding parts of the function $r(\la)$ which
stands at the right-hand side of (\ref{barbaromfunc}). To this end we define
three sums together with some boundary terms
\bea
&& r_1(\la):=\sum\limits_{j\ge 2}\Bigl(\frac{b_{2,0,j}}{(\la-3j)^2}-\frac{b_{2,0,j+1}}{(\la+3j)^2}\Bigr),\nn\\
&& r_2(\la):=\sum\limits_{j\ge 2}\Bigl(
\frac{b_{1,0,j}}{\la-3j}+\frac{b_{1,0,j+1}}{\la+3j}
+\frac{b_{1,1,j}}{\la-1-3j}+\frac{b_{1,1,j+1}}{\la+1+3j}\Bigr)
-\frac{c_{1,0,1}}{\la-6}-\frac{c_{1,0,2}}{\la+3}-
\frac{c_{1,1,1}}{\la-7}-\frac{c_{1,1,2}}{\la+4},\nn\\
&& r_3(\la):=\sum\limits_{j\ge 2}\Bigl(\frac{b_{2,1,j}}{(\la-1-3j)^2}-\frac{b_{2,1,j+1}}{(\la+1+3j)^2}\Bigr)
+\frac{c_{2,1,2}}{(\la+4)^2}-\frac{c_{2,1,1}}{(\la-7)^2},\nn
\ena
which can be calculated explicitly, for example, with the help of
{\it Mathematica}.
With this definition we get the equation
$$
I_0(\la-3)-I_0(\la)=r_1(\la)+r_2(\la)+r_3(\la).
$$
The sum of the functions $r_1(\la),r_2(\la),  r_3(\la)$ is regular 
for $-7<\Re{\la}<4$. Hence, we can write down the above 
first term $I_0$ as an integral
\bea
I_0(\la)&:=&-\int\limits_{-i\infty}^{i\infty} 
\frac{d\mu}{6i}\tan\Bigl({\frac{\pi}{3}(\la-\mu)}\Bigr)S(\mu)\label{I0},
\ena
where the function $S$ in the integrand is
\bea
S(\la)=-S(-\la)=r_1(\la+\frac32)+r_2(\la+\frac32)+r_3(\la+\frac32),
\ena
while the residual term $I_1$ is a rational function of $\la$
\begin{multline}
I_1(\la) := -a\Bigl(\frac1{(\la-1)^2}+\frac1{(\la+1)^2}\Bigr)
+ \Bigl(d+\frac{6}5 a\Bigr)\Bigl(\frac1{(\la-4)^2}+\frac1{(\la+4)^2}\Bigr) \\
-b\Bigl(\frac1{\la-1}-\frac1{\la+1}\Bigr)
-\Bigl(c+\frac3{5}b\Bigr)\Bigl(\frac1{\la-4}-\frac1{\la+4}\Bigr).
\label{I1}
\end{multline}
The function $S(\la)$ in the above definition of the integral $I_0$ can be 
explicitly calculated:
\begin{multline}
{\ds S(\la):=\frac{\la(48\la^4-280\la^2+259)}{2\, d_1(\la)^2}\cdot \Bigl(\psi\Bigl(\frac12+\frac{\la}3\Bigr)+\psi\Bigl(\frac12-\frac{\la}3\Bigr)-
	\psi\Bigl(\frac56+\frac{\la}3\Bigr)-
	\psi\Bigl(\frac56-\frac{\la}3\Bigr)\Bigr)} \\
{\ds + \frac{4 \la^2-9}{48\,d_1(\la)}\cdot \Bigl(\psi_1\Bigl(\frac12+\frac{\la}3\Bigr)-\psi_1\Bigl(\frac12-\frac{\la}3\Bigr)-
	\psi_1\Bigl(\frac56+\frac{\la}3\Bigr)+\psi_1\Bigl(\frac56-\frac{\la}3\Bigr)\Bigr)} \\
{\ds -\frac{2\,\la\,(68\la^2-617)}{5\,d_2(\la)}\cdot \Bigl(\psi_1\Bigl(\frac13\Bigr)-
	\frac{5\pi^2}6\Bigr)+
	\frac{\la\, p_1(\la)}{5 \,d_2(\la)^2}\cdot \Bigl(3\log{3}+\frac{\pi}{\sqrt{3}}\Bigr)+
	\frac{2 \,\la\, p_2(\la)}{375\, d_2(\la)^2}}.
\label{S}
\end{multline}
where
\bea
&& d_1(\la):=(2\la-5)\,(2\la-1)\,(2\la+1)\,(2\la+5),\label{d1d2p1p2}\\
&& d_2(\la):=(2\la-11)\,(2\la-5)\,(2\la-1)\,(2\la+1)\,(2\la+5)\,(2\la+11),\nn\\
&& p_1(\la):=256\, \la^8 - 54528\, \la^6 + 1190496\, \la^4  - 5423440\, \la^2 + 4738305,\nn\\
&& p_2(\la):=18032\, \la^4   - 759448\, \la^2 +  5395775\nn.
\ena
If the argument $\la$  
is purely imaginary, $\la=i x$ with $x\in\Re$, the function $S(\la)$ is purely imaginary as well. 
The  behavior of the imaginary part $\Im{(S(i x))}$  is shown in Fig.~\ref{picture}.
At infinity it approaches 0 as $x^{-3}$. Hence, the integral (\ref{I0})
is convergent. 

It is important that the function $S(\la)$ is regular in the interval 
$-11/2< \Re{\la}< 11/2$. It guarantees that we can shift the integration contour by $\pm 3$  in the 
integral (\ref{I0}) along the real axis without catching any singularities
that would come from the function $S$. There is only one contribution that comes
from the residue at $\mu=\la-3/2$ where the function $\tan(\pi/3(\la-\mu))$
has a pole. It provides a non-trivial result for the 
difference $I_0(\la-3)-I_0(\la)$.
We imply here that the integral with shifted argument $I_0(\la-3)$ must be determined as analytical 
continuation of $I_0(\la)$. It means that together with the shift
of the spectral parameter $\la$ also the integration contour should
be shifted by -3 along the real axis. 

\begin{figure}[h]
	\includegraphics[scale=1.2]{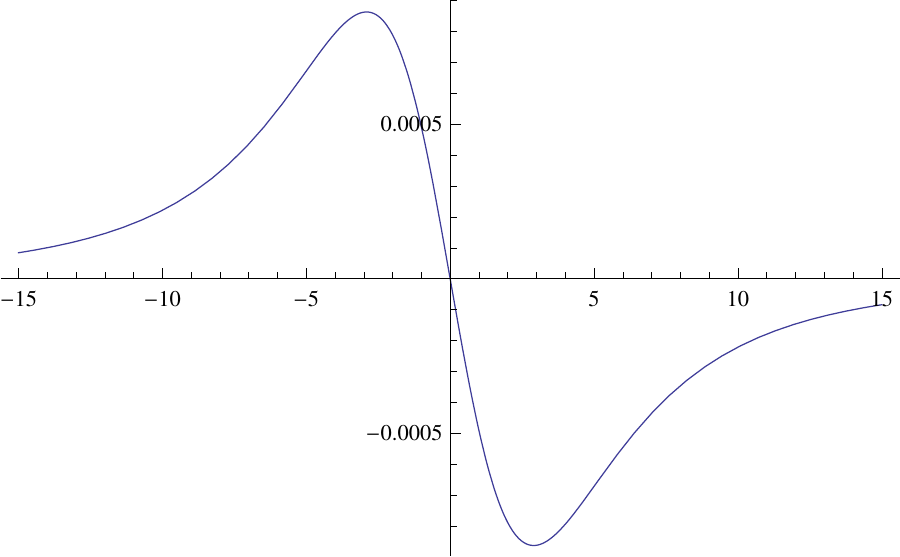}
	\caption{Behavior of the function $\Im (S(i x))$}
	\label{picture}
\end{figure}



Finally, we can check that the solution (\ref{barbarom3res1}) indeed 
satisfies the functional equation (\ref{barbaromfunc}).

Actually, we can explicitly take the integral for the rational part of 
the function $S(\la)$ in (\ref{I0}) and combine it with the term $I_1(\la)$. 
After some algebra we can bring the sum of the terms $I_0$ and $I_1$ to the form (\ref{barbarom3res}). 
Actually, the function $s(\la)$ defined in (\ref{s}) exactly corresponds to the part of the function 
$S(\la)$ (\ref{S}) which depends on the functions $\psi$ and $\psi_1$.

As discussed in Section 4.4, the integral representation (\ref{I0}) that we derived here is useful for 
the study of the asymptotic behavior when the spectral parameter $\la$ is close to $0$. 
Unfortunately, it does not seem very useful for the investigation of asymptotic behavior at large 
$|\la|\to\infty $. Although our numerical study supports the expected behavior (\ref{asympbarbarom}), 
for the moment we cannot approve it analytically. We seem to need some other integral representation. 
So, we leave this question for future consideration.

{\it Acknowledgements.}\quad The authors are grateful to  A. Isaev, 
A. Kl{\"u}mper, A. Razumov, G. Ribeiro and F. Smirnov for many stimulating discussions.  
Our special thanks go to F. G{\"o}hmann for careful reading of the manuscript and
many useful suggestions.
AH is grateful to J.~Sirker and A.~Wei{\ss}e for advises  concerning numerical calculations.

The authors would like to thank {\it Deutsche Forschungsgemeinschaft}
for support within the framework of the DFG {\it Forschergruppe}
FOR 2316, projects number BO 3401/1-1 and GO 825/8-1. The work 
of Kh.S.N. was supported in part by the RFBR grant \# 16-01-00473 
and by the Russian Academic Excellence Project '5-100'. 
		
Note added: After our paper appeared on the arXiv,	we became aware of the related 
preprint \cite{KlumpRib}.

\providecommand{\bysame}{\leavevmode\hbox to3em{\hrulefill}\thinspace}


\end{document}